\documentclass[12pt,onecolumn,showpacs,amssymb,aps,nofootinbib,floatfix]{revtex4-2}
\usepackage{amsmath}
\usepackage{epsfig}
\usepackage{color}
\usepackage{tensor}
\newcommand{\ave}[1]{\left\langle #1 \right\rangle}
\renewcommand{\eqref}[1]{Eq.~(\ref{#1})}
\newcommand{\figref}[1]{Fig.~(\ref{#1})}

\newcommand{\di}{{\rm d}}
\def\wT{{\widehat T}}

\newcommand{\tcomm}[1]{\mathcal{T}\left[ #1\right]}

\newcommand{\eqcomma}{\phantom{AA},\phantom{AA}}

\newcommand{\lnz}{\ln \mathcal{Z}}

\newcommand{\order}[1]{ \mathcal{O} \left( #1 \right) }

\usepackage{soul}

\usepackage[usenames,dvipsnames]{xcolor}
\definecolor{purp}{RGB}{153,50,204}

\begin{document}
	\title{Fluctuating relativistic dissipative hydrodynamics as a gauge theory}
	\author{Travis Dore$^a$,Lorenzo Gavassino$^b$,David Montenegro$^c$,Masoud Shokri$^d$, Giorgio Torrieri$^e$}
	\affiliation{$\phantom{A}^a$Illinois Center for Advanced Studies of the Universe, Department of Physics, University of Illinois at Urbana-Champaign, Urbana, IL 61801, USA\\  $\phantom{A}^b$Nicolaus Copernicus Astronomical Center, Polish Academy of Sciences, ul. Bartycka 18, 00-716 Warsaw, Poland\\  $\phantom{A}^c$Bogoliubov Laboratory for Theoretical Physics, Joint Institute for Nuclear Research, RU-141980 Dubna, Moscow region, Russia\\$\phantom{A}^d$IPM, School of Particles and Accelerators, P.O. Box 19395-5531, Tehran, Iran\\ $\phantom{A}^e$Universidade Estadual de Campinas - Instituto de Fisica "Gleb Wataghin",
		Rua Sérgio Buarque de Holanda, 777,
		CEP 13083-859 - Campinas SP}
	\begin{abstract}
		We argue that different formulations of hydrodynamics are related to uncertainties in the definitions of local thermodynamic and hydrodynamic variables. We show that this ambiguity can be resolved by viewing different formulations of hydrodynamics as particular "gauge choices" which lead to the same physical behavior of the system. Using the example of bulk viscosity, we show that Bemfica-Disconzi-Noronha-Kovtun (BDNK) and Israel-Stewart hydrodynamics are particular "gauge choices" of this type, related by a well-defined transformation of thermodynamic and hydrodynamic variables. We argue that this gauge ambiguity is necessary to ascertain the causality of stochastic hydrodynamic evolution and conjecture that it could explain the applicability of hydrodynamics outside its expected regime of validity since far from equilibrium and close to equilibrium may be related through transformations of this type.
	\end{abstract}
	\maketitle
	
	\section{Introduction}
	Relativistic hydrodynamics has proven to be an immensely successful phenomenological approach, effective for systems as small as hadronic collisions, and as large as the universe as well. Its theoretical origin is in principle clear. It arises as an effective theory for a system with a ``high density of particles'' which also thermalizes on a small scale with respect to the microscopic gradients.
	The deviations from full thermalization could then be incorporated in dissipative coefficients that are associated with gradients of hydrodynamic variables. Nevertheless, a comprehensive account of how hydrodynamics emerges from statistical mechanics is still lacking.
	The relationship between hydrodynamics as a deterministic theory of conservation laws and statistical mechanics, which explicitly relies on microscopic fluctuations, is in tension. Based on statistical mechanics, we may be able to determine ``how far from local equilibrium'' the system has to be before hydrodynamics no longer serves as an approximate description.
	As a consequence of the success of the hydrodynamic description on comparatively small and high gradient systems, such as proton-proton collisions, these issues need to be addressed~\cite{zajc, kodama, crooks}.
	
	Hydrodynamics as an effective theory relies on only conserved currents, whose total flux commutes with the microscopic Hamiltonian~\cite{nicolis,nicolis2,beta,jackiw,glorioso,grozdanov,ryblewski,montenegro}. In the absence of chemical potentials, the only conserved current is the energy-momentum tensor, $T^{\mu\nu}$. In spite of this, the assumption of approximate local equilibration still requires the local rest frame (LRF), which itself requires defining the four-velocity fluid (or fluid flow), which are additional details. This is a result of the ergodic assumption that time averages can be approximated by ensemble averages, a fundamental assumption in statistical mechanics, and the fact that time averaging is frame-dependent under special relativity.  
	Considering fluid dynamics as a coarse-grained description, one divides the medium into space-time domains (fluid cells) for which the macroscopic quantities are approximately constant, but their size is much larger than the microscopic physics scale. In the LRF, a thermodynamic limit is a good approximation within the cell if the fluid is non-dissipative. The LRF represents the comoving frame in each cell. Nevertheless, out of equilibrium, the definition of the fluid four-velocity and the LRF are ambiguous. To remove this ambiguity, one traditionally uses the Landau~\cite{landau} or Eckart~\cite{Eckart} definitions for hydro frame selection\footnote{The reader should be aware of the terminology here. Ref.~\cite{kovfluct} explains hydro frames in detail.}. These two options are identical if the fluid is uncharged, which is the assumption we have made throughout this study. First-order hydrodynamics, the so-called Navier-Stokes (NS) hydrodynamics, is known to be unstable for both choices\footnote{In this work, the term NS is exclusively used for the first-order hydrodynamics in Landau and Eckart frames.}~\cite{hislim,sc-problem-85,sc-problem-87}. In particular, the Landau frame is defined by
	\begin{equation}\label{eq:landau}
		u_\mu T^{\mu \nu}=eu^\nu\eqcomma \Delta^{\mu}_{\alpha}u_\beta T^{\alpha\beta} = 0\,,
	\end{equation}
	wherein $e$ is the equilibrium energy density, $\Delta^{\mu\nu}\equiv g^{\mu\nu} - u^\mu u^\nu$, and $u^\mu$ is the fluid four-velocity. This leads to the transversality condition
	\begin{equation}
          \label{trans}
		u_\mu \Pi^{\mu\nu} = 0
	\end{equation}
	where $\Pi^{\mu\nu}$ contains all out of equilibrium corrections to the energy-momentum tensor.
	With first-order hydrodynamics, one introduces terms of first-order in derivatives of hydrodynamic variables, coupled to new transport coefficients. The resulting energy-momentum tensor can be written as
	\begin{equation}
		\label{eq:gentensor}
		T_{\mu \nu} = (e+p)u_\mu u_\nu +pg_{\mu \nu}+q_\mu u_\nu+q_\nu u_\mu +\left. \Pi_{\mu \nu}\right|_{\propto \partial u}\,,
	\end{equation}
	wherein $q_\mu$ is a vector transverse to $u_\mu$.
	Typically, one uses the equations of motion from the zeroth-order to eliminate the so-called off-shell terms, i.e. terms that vanish when the equations of motion are satisfied. Based on the Kubo formulae, it is possible to derive the transport coefficients from microscopic theory~\cite{jeon}.   At tree-level, these formulae coincide with estimates from the Boltzmann equation.
	
	As of recently, Israel-Stewart (IS) hydrodynamics has been the main approach to link hydrodynamics with far-from-equilibrium systems and remedy the instability of NS hydrodynamics~\cite{IsraelStewart}. In this approach, new dynamical variables are introduced into the nonequilibrium part of the energy-momentum tensor, $\Pi^{\mu\nu}$, which relax to their NS counterparts on a time scale quantified by the theory's relaxation time. Schematically
	\begin{equation}
		\label{iseom-general}
		\tau_X u^\mu \partial_\mu  X + X = X_{\rm NS}\,,
	\end{equation}
	in which $X$ is a hydrodynamic variable expressed in terms of an arbitrary rank tensor.  More complicated, oscillatory behavior is also possible if one adds second-order derivatives in time~\cite{janik}. These theories are usually treated as "top-down effective theories", with the "UV completion" being either a weakly coupled Boltzmann equation or a strongly coupled theory with a gauge dual.   However, we know from condensed matter physics \cite{trachenko} that hydrodynamics emerges from a much wider and richer range of systems, notably systems which are strongly correlated at a short scale and more advective-diffusive at a longer one.
	
	Recently, a new class of theories was developed, the so-called Bemfica-Disconzi-Noronha-Kovtun (BDNK) theory, where $\Pi_{\mu \nu}$ is entirely determined by gradients of the hydrodynamic variables \cite{Bemfica:2017wps,kovtun,kovtun2,bdnk,Bemfica:2020zjp,spal}, based on earlier proposals \cite{yarom,biro,biro2} which reduce the degrees of freedom at the expense of local isotropy. \color{black}
Numerical simulations \cite{bantilan} have confirmed that when the Knudsen number is small, and only in that limit, do the two theories coincide, and this has been interpreted as the statement that while BDNK-type theories might be numerically easier to simulate they would not provide an explanation to the ``unreasonable effectiveness of hydrodynamics``.\color{black}

There are some basic differences between the BDNK first-order theory and traditional NS hydrodynamics. First, in the BDNK theory, the eigenvalues of neither the energy-momentum tensor nor the current correspond to physical quantities since $u_\mu$ is a somewhat arbitrarily chosen vector field. Second, the off-shell terms in the gradient expansion are kept. The price is an uncertain relation between microscopic thermodynamics, entropy, and hydrodynamic evolution since it is unclear what is the right ensemble to sample to get the equation of state and transport coefficients. In addition, the entropy production density $\partial_\mu S^\mu$ is not always positive (sometimes $\partial_\mu S^\mu <0$ for large gradients \cite{gavassino,ShokryTaghi2020,GavaUeit2021}). This last point could be rationalized by assuming that for large gradients the entropy gradient is not the one obtained from a gradient expansion. However, given that fluctuation and dissipation are correlated, transient negative entropy intervals are not necessarily unphysical by themselves, they just reflect stages in evolution where thermodynamic fluctuations are non-negligible.
	
	To properly understand these ambiguities, one must realize that the fluid four-velocity $u_{\mu}$, energy density $e$, fluid pressure $p$, the heat current $q_\mu$, and shear tensor $\pi_{\mu \nu}$ may be considered unobservable figments of the theorists' imagination, since what is actually observed is the entire $T_{\mu \nu}$. 
	The effect hydrodynamic fluctuations have on these dynamics, particularly on $\Pi_{\mu \nu}$, heighten this ambiguity \cite{crooks}.
	Ultimately, experimental observables (such as elliptic flow and higher harmonics) reduce to correlations $\ave{T_{\mu_1 \nu_1}(x_1)...T_{\mu_n \nu_n}(x_n)}$. Part of the approximation of local equilibration is that we cannot tell which ``microscopic particle'' (molecule, quark, gluon) contributes to ``equilibrium'' energy density, pressure, $u_\mu$ and which is ``not thermalized'' and contributes to $\Pi_{\mu \nu}$. 
	
 	The choice of \textit{frame} is, therefore, more similar to ``gauge fixing'' than to a real physical distinction.   As a related point, Fig.~\ref{ambig} illustrates that the relative contribution of thermal fluctuations and thermodynamic evolution to a given observable is experimentally undetectable.   The different ways the two contributions contribute give a physical justification for this ``gauge symmetry''. The choice between a Landau, Eckart, or a more anisotropic frame such as BDNK represent a choice of the role of deterministic dissipative evolution versus thermal fluctuations in how a coarse-grained final state evolved from a similarly coarse-grained initial state.  Microscopically this "choice" can be associated with Bayesian inferencing weighted by Gibbsian entropy \cite{jaynes} over an ensemble of hydrodynamic frames. In terms of macroscopic variables, a range of such frame choices determined by each choice's unique local thermal fluctuations and evolution of conserved quantities should be physically equivalent.

\color{black}
This could, in principle, provide an explanation for the unreasonable effectiveness of hydrodynamics \cite{zajc}, provided the effectiveness of hydrodynamics is not anymore judged by the Knudsen number and averaged gradients (in an ensemble average that breaks down for few particles) but rather by the local applicability of the fluctuation-dissipation theorem, i.e. the idea that dynamics is determined by the number of locally available microstates.      In the limit where interactions are ``strong`` and microscopic particles are strongly correlated (an everyday example is the ``Brazil nut effect" \cite{brazilnut}), it is {\em not} guaranteed that a system with more degrees of freedom achieves a more perfect local thermal equilibrium and close-to-ideal fluid behavior.    As the number of degrees of freedom decrease, fluctuations increase but so do {\em redundant descriptions} which are compatible with local equilibrium but have different Boltzmannian (although not Gibbsian) entropies.    Thus, the chance of finding a description that happens to look like ideal nearly hydrodynamics is greater.\color{black}

\color{black}
The authors are aware of one other hydrodynamic formalism which produces causal and stable evolution, namely the divergence type hydrodynamic equations \cite{LIU1986191,geroch,Aguilar:2017ios}. In this work we will not make many connections with this formalism, other than a possible relation in Sec.\ \ref{functional}. We leave for future work to include the effects of hydrodynamic frame and their associated redundancies into a fluctuating description of divergence type theories \cite{Miron-Granese:2020mbf, Calzetta:1997aj}.

        \color{black}
        The rest of the paper will outline how the frame choice ambiguity outlined above can be seen as a freedom of reparametrization similar to gauge symmetry.  First, in the next section \ref{seczub}, we review Zubarev hydrodynamics and the abiguity of the choice of foliation, reflected in the ambiguity of the flow vector $\beta_\mu=u_\mu/T$. In Sec.\ \ref{reparam} we define the reparametrization linking microscopic fluctuations and dissipative evolution and discuss the details of how it is implemented.   We then devote three subsections to it's physical and microscopic  meaning, respectively in Zubarev hydrodynamics (subsection \ref{justzub}), transport theory (subsection \ref{reptransport} where the connection to the Gibbsian vs. Boltzmannian entroy current is elucidated) and effective theory (subsection \ref{justeft}), concluding with considerations on how the reparametrization symmetry affects the causality of the theory (subsection \ref{repcause})

        Then in Sec.\ \ref{bulkvisc} we investigate the particular case of a bulk viscous fluids.   In subsections \ref{match} and \ref{spectral} we show how, in this case, one may relate BDNK like-theories with Israel-Stewart ones by relabeling heat currents as fluctuation-triggered sound waves (and vice-versa), both considered as ``microscopic'' under a Gibbsian picture.   Section \ref{bulkrep} shows how these relabelings can be seen as a particular case of the reparametrizations described in section \ref{reparam}.
        Finally, in Sec.\ \ref{functional}, we explore how the symmetries we consider appear as quasi-conserved currents in the effective action approach to hydrodynamics pioneered in \cite{grossi}\color{black}

	Throughout the paper we adopt the metric signature $(+,-,-,-)$ and work in natural units $c=k_B=\hbar=1$.
	
	\section{Zubarev hydrodynamics and the entropy current \label{seczub}}
\color{black}Hydrodynamics is usually thought of as an effective theory.  This begs the question of what the ``fundamental theory'' is.   In the previous section, we cited Gauge/gravity duality and the Boltzmann equation.   The latter is obviously more directly connected to physically realistic systems, and indeed most derivations of hydrodynamics employ it.   In some approaches, this has been done in the ``Wilsonian'' renormalization group sense \cite{kunihiro,kunihiro2}.   In this respect, anisotropic frames until now were defined in terms of relaxation time matching conditions of the Boltzmann collision term \cite{kovtunani,denicol} (an exact renormalization group treatment requires a frame isotropic w.r.t. entropy \cite{kunihiro2}).
Such derivations however are incomplete since the Boltzmann equation is itself incomplete as a theory.  There are fluids, such as water, where hydrodynamics itself provides a better description than the Boltzmann equation.   Moreover the "best" result that shows that the Boltzmann equation is a good approximation for {\em any} system (in other words that microscopic reversible correlations are irrelevant) is Lanford's theorem \cite{lanford}, a statement of the time-scale of applicability that notably fails to give a long-time limit for hightly correlated systems.

A different way for hydrodynamics, in particular it's generic frame form, to emerge is therefore desireable.
        The most universal "UV theory" might be the fundamental principle of statistical mechanics, the equal distribution of microstates.   As this section and the next one make clear, some care is needed to define the theory this way if the system is not in perfect local equilibrium, i.e. if microstates are not quite equally probable.  We shall argue in the next section that the fluctuation-dissipation theorem, promoted to a "symmetry" can be used in this regime, and anisotropic hydrodynamics emerges as a byproduct of this symmetry.    Here, we shall give a brief introduction to Zubarev hydrodynamics, which has the advantage of being defined ab initio via equilibrium and microstate equality. 
	Fundamentally, equilibrium is defined in terms of time-symmetric microscopic states, via the KMS condition \cite{kadanoff}, occupied uniformly under the constraint of conservation laws.\color{black}
	
	In the coarse-graining of the fluid one divides the medium into spacetime domains, the so-called fluid cells, for which the macroscopic quantities are approximately constant, but their size is much larger than the microscopic physics scale. If the fluid is non-dissipative, then for the comoving observer, the thermodynamic limit\footnote{Intended as extensivity, rather than as limit where fluctuations are negligible; this paper applies in the situation where they are not negligible with respect to the mean free path} is a good approximation within the cell. Hence such an observer can in principle derive the macroscopic observables from an underlying microscopic theory. Such an underlying theory in the most fundamental level may be encoded in a Lagrangian and therefore the usual methods of thermal field theory can be employed. For example (we neglect the chemical potential throughout this work for simplicity, and therefore Landau and Eckart frames are identical):
	\begin{equation}
		\label{micro}
		e=-\frac{\delta}{\delta \beta}\lnz \eqcomma p=\beta^{-1}\lnz\,. 
	\end{equation}
	Here $\mathcal{Z}$ is the partition function, and $\lnz$ is obtainable from a functional integral of the microscopic Lagrangian\footnote{The equilibrium part being defined using the Matsubara time prescription, which stipulates that time is periodic}~\cite{gale,huang}
	\begin{equation}
		\label{lnzeq}
		\mathcal{Z} = \int \mathcal{D}\phi \exp[S[\phi]]\,,
	\end{equation}
	wherein $S[\phi]$ is the Euclidean periodic action. The Landau frame is special in the sense that \eqref{eq:landau} continues to hold for the dissipative fluid.
	%
	In the grand canonical ensemble, neglecting quantum corrections and correlations as well as chemical potentials for simplicity, we have 
	\begin{equation}
		\label{cluster}
		\mathcal{Z}= \sum_{i=0}^\infty  \mathcal{Z}_N \eqcomma \mathcal{Z}_N = \frac{1}{N!} \int \prod_{j=1}^N d^3p_j \, d^3 x_j \exp\left[ -\frac{E_i(p_1,...,p_N)}{T}  \right] 
	\end{equation}
	The Gibbs-Duhem relation
	\begin{equation}
		s\equiv \frac{dp}{dT}=\frac{p+e}{T} \eqcomma T=\beta^{-1} 
	\end{equation} 
	follows automatically from \eqref{micro} and guarantees that the equilibrium entropy current is conserved, $\partial_\mu \left( s u^\mu\right)=0$. 
	The above definitions are quite basic, in that they directly follow from the basic assumptions of statistical mechanics.
	In that respect, in equilibrium LRF is ``special''.
	However, 
	this frame can only be unambiguously known if ``neighboring cells'' used to coarse-grain the fluid (separated by a sound-wave dissipation length) are also each in the thermodynamic limit. This is necessary so that thermal fluctuations factor out, otherwise neighboring cells would drive the cell of interest out of equilibrium which is a manifestation of the fluctuation-dissipation theorem. But this assumption is stronger than the assumption that the Knudsen number \footnote{Knudsen number is the ratio of particles mean free path to the typical length scale for which the hydrodynamic variables change.} is small, and can not hold for small systems even if gradients/Knudsen numbers are small. On the other hand, those fluctuation-dissipation relations (on which linear response theory is based) are independent of ensemble choice \cite{ensemble1,ensemble2,ensemble3} and so local equilibration could occur without the necessity of assuming that every cell has to be large enough to approach the thermodynamic limit\footnote{Generally \cite{ensemble1}, \eqref{cluster} can be moved to a microcanonical ensemble via \[\ \Omega(E)=\int dT \mathcal{Z} (T) \exp\left[iTE \right]  \]
		equations \ref{eq:transportkubo} will be updated straight-forwardly via linear dispersion relations \cite{ensemble3}.  Temperature fluctuations $\sim T^2/c_V$ while energy fluctuations $\sim c_V/T^2$, where $c_V$ is the second derivative of Eq. \ref{cluster} w.r.t. T~\cite{huang}.}.
	While, since cell coarse-graining is arbitrary, one expects the grand canonical ensemble to apply for fluids, changing between ensembles is in principle straight-forward \cite{ensemble1,ensemble2}, one can also appropriately convert \eqref{cluster} if necessary.

        \color{black}
        Given a field configuration close to thermal equilibrium and a contour respecting energy-positivity and the KMS condition, 
	transport coefficients can be obtained from infrared limits, i.e. $k\to 0$, of the \eqref{eq:correlator} of the energy-momentum tensor \cite{kadanoff,tong,jeon,jeonb,sonreview}, \color{black} using the fluctuation-dissipation relation
	\begin{equation}
		\label{eq:transportkubo}
		\eta \sim \lim_{k \rightarrow 0} \mathrm{Im} \frac{1}{k} \tilde{G}(k) \eqcomma \tau_\pi= \lim_{k\rightarrow 0}\frac{k}{2\mathrm{Im}[G(k)]} \left[ \frac{1}{k^2} - \frac{1}{w(k)^2} \right]\mathrm{Re}[G'(k)]\,,
	\end{equation}
	wherein \color{black}
	\begin{equation}
		\label{eq:correlator}
		\tilde{G}(k)=\int \left. \theta\left(\mathrm{Im}\left(x^0-x^{0'}\right)\right)  e^{ik(x-x')}\frac{1}{g} \frac{\delta^2}{\delta g_{ij}^+(x) \delta g_{ij}^-(x') } \lnz\,. \color{black} \right|_{g_{\mu \nu}=\eta^\pm_{\mu \nu'}} 
	\end{equation}
        and \[\  \eta^\pm_{\mu \nu}= diag\left( \frac{1}{T}\exp\left[ \frac{i2\pi \theta}{T}\right] \pm i\epsilon ,-1,-1,-1 \right) \] is the Minkowski metric with a periodic boundary condition enforcing Matsubara time.
        \color{black}
	Note that all microscopic interactions in the above are in principle calculable via $E(p_i,x_i)$, the Hamiltonian.
	
	However, we note a tension between the development of ``macroscopic causal'' hydrodynamics described in the previous section, and the considerations of statistical mechanics done here.
	The fact that in \eqref{eq:gentensor} the ``LRF'' parameters depend on $u_\mu$  means that the very definition of the equation of state and transport coefficients might be not-trivial, although it might be irrelevant for microscopic evolutions in certain limits \cite{hislim,lindblom}. Even in the Landau frame, where we expect \eqref{eq:transportkubo} to apply straight-forwardly, the positivity of the entropy production relies on the assumption that the ratio $\tau_\pi/(T\eta)$ is the second-order expansion coefficient of the entropy density in terms of $\Pi_{\mu \nu}$, and hence is a merely thermodynamic (not kinetic) quantity~\cite{hislim,lindblom}, \color{black} arising from the pole structure of thermodynamic correlators \cite{laurent} \color{black}.
	However, as \cite{jeon,dirkis} and earlier works show, the relaxation time can be seen as a transport coefficient, given by a Kubo formula. 
	The tension between these pictures reflects the fact that in an interacting theory ``thermodynamic'' and ``transport'' quantities cannot be straightforwardly unentangled.  
	This is the basic reason why transport perturbative calculations exhibit divergences even in weakly coupled theory~\cite{arnold} and also reflects the fact that ``local thermodynamic limit'' and ``small Knudsen number'' refer to different limits and thus need to be handled consistently in a hydrodynamic treatment.   \color{black}These ambiguities are swept under the carpet in the Knudsen number expansion because it is implicitly assumed the gradient is large enough w.r.t. the scale setting the thermodynamic limit that $k\to 0$ in Kubo formulae can consistently render fluctuations irrelevant.  This is {\em not} true in systems of $\order{50}$ particles. \color{black}
	
	As a related issue, strict positivity of the entropy production only holds ``on average'' over ``large distances''~\cite{crooks}.  Beyond this, one has a fluctuation-dissipation theorem~\cite{kadanoff,tong}, which is more directly related to unitarity and time-reversal invariance of the underlying microscopic theory~\cite{tong}. Entropy increase and its maximization in equilibrium should emerge in the ``far-infrared hydrodynamic scale'' w.r.t. microscopic fluctuations.
	However, turbulence and anomalous dissipation~\cite{zeroth} make it likely that microscopic fluctuations back-reacting on hydrodynamic modes will modify infrared transport coefficients (as corrections from interacting sound waves generally do \cite{stick,kovfluct,hnatic}), and the fact that this must be done in a way that maintains causality and stability for {\em all possible effective theories} suggest microscopic-macroscopic correlations need to be systematically understood, and also that such correlations mean equivalent descriptions exist where ``sound wave backreaction'' are replaced by ``thermal fluctuations'' (and vice-versa).

	Because of the fact that any interaction potential will make Gibbsian and Boltzmannian entropy different \cite{jaynes}, the above ambiguity will make the process of equilibration ambiguous, because one could interpret inhomogeneities as correlations present in the partition function or collective excitations that will dissipate and be reseeded as the system evolves.   Deterministic hydrodynamics however neglects this, since one usually assumes Boltzmannian entropy current to be the physically relevant one.
	
	Zubarev hydrodynamics makes some of the issues above clearer: 
	if the system is approximately in local equilibrium, one can 
	\cite{crooks,zubbec} \color{black} impose the KMS condition and assume quantum interference is ``irrelevant'',i.e. it decoheres much faster than the macroscopic evolution.   The first assumption means that the partition function depends not on the microscopic lagrangian but on the energy-momentum tensor $T_{\mu \nu}$ contracted with a field of Lagrange multiplies $\beta^\mu = u^\mu/T$ and a spacetime foliation d$\Sigma_\mu$.
        The second assumption means that the foliation $d\Sigma_\mu$ can be trivially analytically continued.   The partition function is then
        	\begin{equation}
		\label{zuberrho}
		\mathcal{Z}_E(\Sigma,\beta_\mu) = \int \mathcal{D}\phi \exp \left[ - \int_{\Sigma(\tau)} d \Sigma_\mu \beta_\nu \hat{T}^{\mu \nu}    \right]\end{equation}
Since the density matrix is determined by the partition function and boundary conditions \cite{nishioka}, it is easy to see that, at least if one neglects boundary effects, \eqref{zuberrho} produces a density matrix stationary in the frame co-moving with $\beta_\mu$ provided the system was in perfect equilibrium at the first foliation.  This requirement of course can not be generally fulfilled.
      However, if this requirement is fulfilled approximatly using Stoke's theorem \cite{zubbec}, one can expand the density matrix to the first order around the local equilibrium \cite{zubbec}
        \begin{equation}
          	\mathcal{Z}_E(\Sigma,\beta_\mu)
		\simeq \int \mathcal{D}\phi  \exp\left[ - \int_{\Sigma(\tau+d\tau)}  d \Sigma_\mu \;  T^{\mu\nu}
		\beta_\nu + \int_{\partial\Sigma} d\Omega \;  T^{\mu\nu}
		\nabla_\mu \beta_\nu   \right]
		\label{zuberrho2}
	\end{equation}
	where $\partial\Sigma$ is a small rectangle delimiting the foliation $\Sigma(\tau)$ and $\Sigma(\tau+d\tau)$ and $d\Omega$ a surface element of that rectangle. \color{black}One can see by counting gradients that dissipative dynamics comes from the second term only \cite{zubbec}\color{black}

      Since the partition function and the density matrix contain the same information\cite{nishioka}, one would expect to recover the same information of Eq. \eqref{eq:transportkubo} using Zubarev techniques.   Indeed, eq \eqref{zuberrho2} was used in  ~\cite{zubbec} to expand around equilibrium for any operator
	\begin{equation}
		\label{eom}
		\langle \widehat O (x) \rangle - \langle \widehat O (x) \rangle_{\rm LE}
		\simeq i T \int_{t_0}^{t} \!\!\! \di^4 x'
		\int_{t_0}^{t'} \!\!\! \di \theta \; \left(\langle
		[\widehat O(x),\wT^{\mu\nu}(\theta,{\bf x}')] \rangle_{\beta(x)} \partial_\mu
		\beta_\nu(x')  \right)
	\end{equation}
	where $\theta$ refers to the standard imaginary Matsubara time.
	Using these equations, with $\hat{O}\equiv \hat{T}_{\mu \nu}$, the (non-causal) NS equations can be derived from a gradient expansion, as can the Kubo formulae \eqref{eq:transportkubo} for $\eta$ and $\zeta$, as can be seen in \cite{zubbec} which coincide with the metric formulas used in \cite{sonreview}.  Recently this approach was extended to second order coefficients \cite{dirkis}.\color{black}

        \color{black}
        We note that the above formulae, \eqref{zuberrho} and \eqref{eom} are valid only at a scale where thermal fluctuations can be discounted.    One can see this via the gravitational Ward identiy,
       \cite{boulware,jeon,jeon2}
        	\begin{equation}
		\label{ward}
		\partial^\alpha \left\{ \ave{\left[ \hat{T}_{\mu \nu}(x),\hat{T}_{\alpha \beta} (x') \right]} - 
		\delta(x-x') \left( g_{\beta \mu} \ave{\hat{T}_{\alpha \nu}(x')} + g_{\beta \nu} \ave{\hat{T}_{\alpha \mu}(x')}  
		- g_{\beta \alpha} \ave{\hat{T}_{\mu \nu}(x')}  \right) \right\} =0\,,
	\end{equation}
This identity holds for the ``UV'' theory (when all correlations are included) at all scales \cite{jeon2} as a consequence of local Lorentz invariance.    Provided locally a hydrostatic limit holds and in the infrared limit, the ``contact terms'' $\sim \ave{T_{\mu \nu}}$ disappear, and the clean separation in \eqref{zuberrho2} and \eqref{eom} can be recovered.     When the mean free path and the thermal fluctuation scale are comparable and the background is far from hydrostatic, however, they are present and the two terms of \eqref{zuberrho2} \color{black}do not separate, and a clear distinction between ``dissipative'' and ``non-dissipative'' terms becomes impossible (physically, thermal fluctuations mix dissipative and non-dissipative contributions)\color{black}.        

\color{black}
	We can now examine the entropy current issue discussed in the previous section in a bit more detail. As long as the Gibbs stability criterion is respected \cite{gavagibbs}, this non-positivity can be understood by either evoking thermodynamic fluctuations \cite{crooks} or by altering the form of $\Pi_{\alpha \beta}$ \cite{gavassino}.
	From the point of view of Zubarev hydrodynamics \cite{zubbec} this can be understood by examining the entropy current's dependence on the chosen foliation
	\begin{equation}
		\label{noneqent}
		\frac{d}{d\tau} \left( s^\mu d \Sigma_\mu \right) = \frac{1 }{s}s^\mu d \Sigma_\mu\Pi_{\alpha \gamma} \partial^\alpha \beta^\gamma \underbrace{ \geq 0}_{\rm if\phantom{A}2nd\phantom{A}law}\,,
	\end{equation}
	where $d \Sigma_\mu$ is a small 3-volume element (with unit normal $n_\mu$), $\tau$ is the proper time w.r.t. that foliation and $s^\mu$ the entropy current. 
	
	The non-positivity of the entropy production can be readily understood: if one integrated over $d\Sigma_\mu$ (as in \cite{zubbec}), entropy would certainly increase, modulo fluctuations going to zero in the infinite volume.
	But, if $n_\mu$ has nothing to do with the entropy current ($u_\mu$ in the Landau frame), the foliation is sampling ``many cells'' rather than tracking the evolution of a single cell.  The fluctuation-dissipation effective field theory picture suggested by Fig. \ref{ambig} suggests these different pictures are linked, in that a variation of $n_\mu$ w.r.t. the Landau frame $u_\mu$ will lead to decreases in entropy given by a fluctuation-dissipation relation \cite{crooks}.
	Thus, the expansion in \cite{kovtun,kovtun2,bdnk} of \eqref{eq:gentensor} where $q_\mu$, $e$, and $p$ in gradients of an arbitrary flow field $u_\mu$ under the constraints of the second law is equivalent to a choice of $d\Sigma_\mu$ of \cite{zubbec}.    
	
	Indeed, in Zubarev hydrodynamics $d\Sigma_\mu$ is not necessarily parallel to $u_\mu$ \footnote{In fact, if $u_\mu$ is vorticose it cannot be globally parallel to $u_\mu$}. Moving away from a parallel frame is equivalent to introducing anisotropy, and to an extent (as long as one loses all touch with equilibrium) it is a theoretical choice, equivalent to redefining $u_\mu$ and $\Pi_{\mu \nu}$.
	In the next section, we will show how to understand the different reparametrizations in terms of symmetries.
	\section{Reparametrization via symmetries}\label{reparam}

        \color{black}
        As we saw hydrodynamics is usually thought of as an effective theory obtained by coarse-graining a microscopic theory in configurations close to local equilibrium.  Thus
        the usual justification for for choosing different matching conditions \cite{bdnk} is one of ``field redefinitions'' \cite{kovtunani}, common to zero temperature effective field theories.    This justification must however be be taken with some care as it clashes with fundamental principles of statistical mechanics, such as ergodicity.    To clarify the role of redefinitions, thereforre, one must remember that Wilsonian renormalization has a probabilistic interpretation related to the central limit theorem \cite{jlreno}.   In this sense, undetectable correlations and fluctuations at or beyond the cut-off scale in effective field theories are ``washed out'' in the same way that higher cumulants are irrelevant when the sum of many uncorrelated variables is considered.    What has recently become appreciated is that for effective theories with UV completion, field redefinitions bring non-trivial correlations sensitive to UV completion (see for instance \cite{coveft}).   In the language of \cite{jlreno}, transforming the variables before summing them over can affect how close is the central limit reached.
        
        At finite temperature, for fluids with finite viscosity, these microscopic correlations are complicated by the presence of an entropy current which is approximately, but not quite, locally maximized and does not qualify as a conserved current.      \cite{bdnk,kovtunani} formulate the theory in terms of a gradient expansion of conserved quantities, with entropy used as a stability check.    This is similar, but not quite the same, as building an effective theory around fluctuation-dissipation from a local equilibrium state, as the local decreases of entropy pointed out in \cite{gavassino} show.    As we argue in this work, in fact, the latter approach can also provide a justification for anisotropic hydrodynamics, and in addition clarify the role of microscopic correlations in field redefinitions and a justification of the applicability of hydrodynamics generally different from the smallness of the Knudsen number.
        \color{black}
        
\begin{figure}[ht]
          	\includegraphics[width=0.95\textwidth]{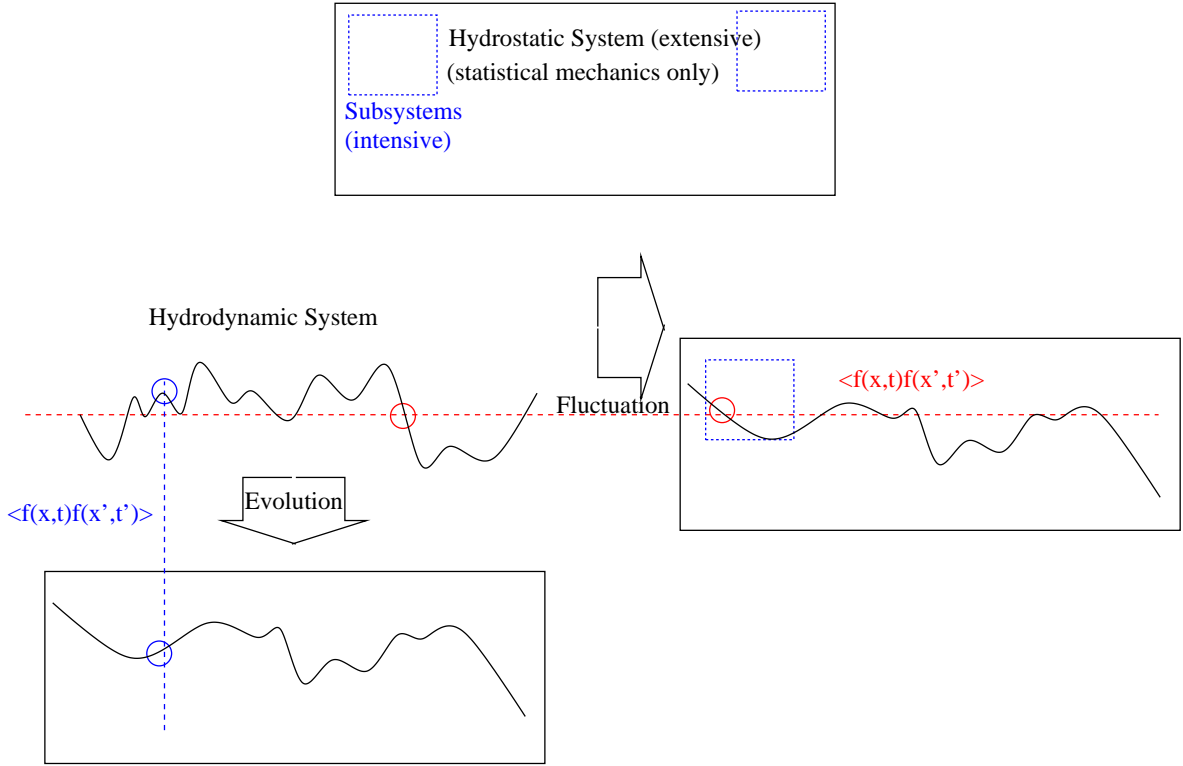}
		\caption{\label{ambig}Ambiguity in a system evolving around local equilibrium.
			A hydrostatic system is essentially what we study within statistical mechanics. In the thermodynamic limit, it can be described either through sub-systems with fluctuating energies, or, equivalently, of fluctuating temperatures.
			If the system is evolving, but approximately in local equilibrium, one must be able to maintain these multiple description choices. This can lead to a variety of equivalent flow frames, where a given correlation in coarse-grained quantities (Here called $f$, which can be $T_{\mu \nu},u_\mu,$ energy density,...) could be given by a microscopic fluctuation or a macroscopic dissipative evolution.  As long as the probability of the two cases is comparable, it means the frame at rest with the flow need not be isotropic.}
	\end{figure}

	A fluid is defined by its symmetries \cite{varnold,jackiw,nicolis,nicolis2,glorioso}, which in the ideal case are homogeneity, isotropy (as passive transformations: they reparameterize the solution), and volume-dependent diffeomorphisms (as active ones, they turn the solution into another one). Non-ideal fluids break some of these symmetries, but a reparametrization symmetry, evident in the freedom of changing between the BDNK and IS approach, is added. As we shall argue in this section,  physically, this symmetry is simply the fact that a correlation in energy-momentum measured at a distance could be the result of a microscopic thermal fluctuation or a macroscopic collective excitation. Measuring this correlation individually gives us no way of knowing its source, and hence many different dynamical configurations could have given rise to it (Fig. \ref{ambig}).
	Within the Zubarev approach, this symmetry reflects the freedom to choose a foliation. One can imagine a ``more time-like foliation'' would make deterministic evolution more prominent, and a ``more space-like foliation'' would enhance thermal fluctuations. But since foliation is a reparametrization symmetry, actual physical observables would not be touched.
	Here we try to understand this symmetry from a Lorentz group theory point of view.   In general, for the energy-momentum tensor to be decomposed as \eqref{eq:gentensor}, one must set up a flow $dM_{\mu \nu}$ (which is just a Lorentz boost)\footnote{Note that this discussion changes drastically in the non-relativistic limit, since the group of symmetries there is not a subgroup of the Lorentz group.  See last paragraph of section \ref{bulkrep}}
	\begin{equation}
		\label{flowstep}
		\Pi_{\mu \nu} \rightarrow \Pi_{\alpha \gamma} \left( g^\alpha_\mu g^\gamma_\nu- g^\alpha_\mu d M^\gamma_{ \nu} - g^\gamma_\nu d M^\alpha_{ \mu} \right) \eqcomma u_\mu \rightarrow u_\alpha \left( g^\alpha_\mu- dM^\alpha_\mu  \right)
	\end{equation}
	as well as the equation of state \eqref{lnzeq}
	\begin{equation}
		\label{bulktrans}
		\lnz \rightarrow \lnz + d\lnz   
	\end{equation}
	\[\ d\lnz =\sum_{N=0}^\infty \int \prod_{j=1}^N d^4 p_j \delta\left( E_N(p_1,...p_j)-\sum_j p_j^0 \right)  \sqrt{\mathrm{det} \left[dM\right]} \exp \left(- \frac{ dM_{0\mu}p^\mu }{T} \right)  \]
	Note that it is only the $d^3 p$ integral which is Lorentz transformed at the microscopic level. Instead, the coarse-grained volume interval is just boosted. This is because the volume element Lorentz transformation is taken care of by \eqref{flowstep}. Note that this is consistent with the separate (but related) interpretation of $M_{\mu \nu}$ as inducing a thermodynamic fluctuation w.r.t. a flow frame defined by the Landau condition \eqref{eq:landau}.
	In terms of the ergodic hypothesis, mentioned earlier, these transformations amount to "biasing" the local comoving phase space sampling together with a deformation of the local time, to model a local fluctuation in the equilibrium frame.
	For ideal hydro with negligible fluctuations (large heat capacity), this separation would be impossible since equilibrium phase space is a Lorentz scalar and $\Pi_{\mu \nu}=0$ in \eqref{flowstep}.
	
	\eqref{bulktrans} is unwieldy (and the factoring out of $dM_{\mu\nu}$ is even more unwieldy, which is why we left the equation in this form) but just reflects equilibrium statistical mechanics in a transformed frame.  It is necessary because this transformation is just a reparameterization, leaving the total $\ave{T_{\mu \nu}}$ invariant. Equations (\ref{bulktrans}) and (\ref{flowstep}) contribute to this, forming a constraint
	\begin{equation}
		\label{avconst}
		dM_{\mu \nu}(x) \frac{\delta \lnz_E[\beta_\mu]}{d g^{\alpha \mu}(x)}=0 
	\end{equation}
	where $\lnz_E$ is given by \eqref{zuberrho}. This constraint is nothing but a non-linear generalization of the Gibbs-Duhem relation \cite{Israel2009,IsraelStewart,gavagibbs}
	\begin{equation}
		\label{israelgibbs}
		\delta s^\mu = - \beta \mathcal{K}_\nu \delta T_0^{\mu \nu} = \beta \mathcal{K}_\nu \delta \Pi^{\mu \nu} \eqcomma \mathcal{K}_\nu M^{\mu \nu} \propto \mathcal{K}^\mu\,.
	\end{equation}
	where $\mathcal{K}_\mu$ is a killing vector of the co-moving frame ($\ne u_\mu$ away from ideal hydrodynamics).
	Since transformations defined by $M_{\mu \nu}$ correspond to different foliations they can in principle change entropy and the equilibrium part of the energy-momentum tensor $T_0^{\mu \nu}$, but the relation between the two must be invariant. 
	
	Because the Boltzmann entropy need not be invariant under parametrizations, the Gibbs stability criterion \cite{Israel2009} ($\delta^1 (s_\mu+ \delta \beta^\nu T_{\mu \nu}=0,\delta^2(...)\leq 0$) becomes probabilistic (configurations that recpect it are {\em likely} to evolve hydrodynamically, without fluctuations affecting it, while configurations that violate it are {\em likely} to be fluctuations-dominated).
	Thus, as long as such transformations maintain the reference frame close to a comoving equilibrium frame, they are unlikely to violate the Gibbs criterion, with probabilities $P(....)$ of entropy decreasing weighted as 
	\begin{equation}
		\label{crooks}
		\frac{P\left( T_0^{\mu \nu}\rightarrow T_0^{\mu \nu}+\delta T_0^{\mu \nu} \right)}{P\left(  T_0^{\mu \nu}+\delta T_0^{\mu \nu} \rightarrow  T_0^{\mu \nu}\right)}
		\sim \exp\left[ \mathcal{K}_\mu \delta s^\mu \right] 
	\end{equation}
	in analogy with \cite{crooks}.
	Equation \eqref{israelgibbs} together with the Boltzmann interpretation of entropy implement this condition, which can be taken as a good definition of ``close to local equilibrium''.
	
	Since $\lnz_E$ depends on $\beta_\mu$ only, it is clear that these transformations are defined in the comoving frame, so one can automatically satisfy \eqref{avconst} up to an overall constant by having $dM_{\mu \nu}$ of the form
	\begin{equation}
		\label{element1}
		dM_{\mu \nu} = \left(\Lambda \right)^{-1}_{\alpha \mu} \Lambda_{\beta \nu} dU^{\alpha \beta}\,,  
	\end{equation}
\color{black}
where the ``kernel'' part 
$dU^{\alpha \beta}$ performs the infinitesimal replacement of \eqref{flowstep}.  Since, as argued earlier, possible changes can occur in any component of $\Pi_{\mu \nu}$ generators involving all components will be necessary:
	\begin{equation}
		\label{element2}
		dU^{\mu \nu} = \eta^{\mu \nu} d\mathcal{A}+ \sum_{I=1,3} \left( d\alpha_I \hat{J}^{\mu \nu}_I + d\beta_I \hat{K}^{\mu \nu}_I \right)
	\end{equation}
	The first term is the bulk channel, the rest are the shear ones, namely
	\begin{equation}
		\label{kproj}
		K_1 = \left(
		\begin{array}{cccc}
			0 & 1 & 0 & 0\\
			1 & 0 & 0 & 0\\
			0 & 0 & 0 & 0\\
			0 & 0 & 0 & 0\\
		\end{array}
		\right) \eqcomma K_2=  \left(
		\begin{array}{cccc}
			0 & 0 & 1 & 0\\
			0 & 0 & 0 & 0\\
			1 & 0 & 0 & 0\\
			0 & 0 & 0 & 0\\
		\end{array}
		\right) \eqcomma K_3=  \left(
		\begin{array}{cccc}
			0 & 0 & 0 & 1\\
			0 & 0 & 0 & 0\\
			0 & 0 & 0 & 0\\
			1 & 0 & 0 & 0\\
		\end{array}
		\right)  \end{equation}
	which move components from $\Pi_{\mu \nu}$ to $q_\mu$ (see \eqref{eq:gentensor})  as well as $J_{1,2,3}$
	\begin{equation}
		\label{jproj}
		J_1 = \left(
		\begin{array}{cccc}
			0 & 0 & 0 & 0\\
			0 & 0 & 1 & 0\\
			0 & 1 & 0 & 0\\
			0 & 0 & 0 & 0\\
		\end{array}
		\right) \eqcomma J_2=  \left(
		\begin{array}{cccc}
			0 & 0 & 0 & 0\\
			0 & 0 & 0 & 1\\
			0 & 0 & 0 & 0\\
			0 & 1 & 0 & 0\\
		\end{array}
		\right) \eqcomma J_3=  \left(
		\begin{array}{cccc}
			0 & 0 & 0 & 0\\
			0 & 0 & 0 & 0\\
			0 & 0 & 0 & 1\\
			0 & 0 & 1 & 0\\
		\end{array}
		\right)  \end{equation}
        Note that these transformations generally break the transversality condition \eqref{trans}.
        Note also that these generators are not arbitrary, because not all combinations solve \eqref{flowstep}, \eqref{bulktrans} and \eqref{avconst}.
Because of the non-linearity of the $\Lambda \Lambda^{-1}$ step in \eqref{element1}, mapping out allowed valueds of the coefficients is generally not possible, similarly to a non-Abelian gauge constraint equation.\color{black}
               
	For any starting $\Pi_{\mu \nu}$ equation \eqref{renoflow} can be integrated via a flow parameter $s$
	\begin{equation}
		\label{renoflow}
		\Pi_{\mu \nu}' = \int_{s}  \left[ \eta^{\mu \nu} \frac{d\mathcal{ A}}{ds}+ \sum_{I=1,3} \left( \frac{d \alpha_I}{ds} \hat{J}^{\mu \nu}_I + \frac{d \beta_I}{ds} \hat{K}^{\mu \nu}_I \right) \right]\Pi_{\mu \nu} ds
	\end{equation}
	A fixed point in this evolution, where $\Pi_{\mu \nu}$ is a function of equilibrium quantities and their gradients indicates the transformation of the initial 
conditions into a BDNK-like ``Gauge''.
By symmetry this will force $\Pi_{\mu \nu}$ to be a function of the gradient and an 
an integration constant vector (identified with the heat flow $q_\mu$), recovering \eqref{eq:gentensor}
\[\ \Pi_{\mu \nu} \rightarrow q_\mu u_\nu + q_\nu u_\mu + f(u,\partial u)  \]
	A comparison with \eqref{noneqent} as well as \eqref{israelgibbs} shows that to first order the entropy production at rest with the foliation, 
	$\sim \Pi^{\mu \nu} \partial_\mu u_\nu$ is zero.  Changing $\Pi_{\mu \nu}$ also changes $u_\mu$ in a way that that balances out the total energy momentum and the entropy, the latter because the energy taken out of $\Pi_{\mu \nu}$ goes into a "sound wave created by the thermal fluctuation".  Because of the overall invariance of $T_{\mu \nu}$ (\eqref{avconst}) these necessarily balance.   However, beyond the linearized limit, sound waves will interact, changing that entropy. This change reflects the fact that outside of the ideal fluid limit Gibbsian and Boltzmannian entropy (the latter identified via \eqref{noneqent}) are not the same \cite{jaynes}.  Gibbsian entropy, defined to also include sound-wave fluctuations and interactions lost in the coarse-graining, is {\em defined} as any entropy missing from the Boltzmannian description.   The next two sections give possible ways to understand this entropy quantitatively, in terms of both microscopic and macroscopic perturbations	
	\subsection{Reparametrization and the Zubarev ensemble \label{justzub}}
	\color{black}
	Within the Zubarev approach, the link between the reparametrization advocated here and "real" gauge theories is particularly apparent.
We shall start with a very short introduction to the role of gauge symmetry in quantum and statistical field theory \cite{peskin}, and then compare with Zubarev hydrodynamics.
         
Fundamentally a Gauge theory arises from the fact that there is a redundancy in the functional integral, since the integral is over some vector field $A_\mu$ and the action only depends on the field strength $F_{\mu \nu}$\footnote{The ambiguity of the contour distinguishing between real and imaginary time, discussed above equation \eqref{zuberrho} and below \eqref{eq:correlator} holds here. $S$ could be in real or imaginary time.}
\[\
\mathcal{Z} = \int \mathcal{D} A^{\mu} \exp\left[ S[F_{\mu \nu} \right] \equiv \int \mathcal{D} A^{\mu}_1 \mathcal{D} A^{\mu}_2 \exp\left[ S[A^\mu_1 \right]     \]
here $A^\mu_{1,2}$ are respectively Gauge equivalent and non-equivalent configurations.     These two classes of field configurations cannot generally be cleanly separated but we remember that physics is not a function of $\mathcal{Z}$ but rather of derivatives of $\lnz$.   Hence, one can choose a gauge $G(A_\mu)=0$ and, up to a in finite but physically irrelevant constant $\Lambda$ we have
\[\   \lnz = \Lambda + \lnz_G \eqcomma Z_G = \int \mathcal{D} \mathcal{A^\mu} \delta \left(   G(A^\mu) \right) \exp \left[ S(A_\mu) \right] \]
this decomposition is easier to handle, although the $\delta-$term causes spurious "ghost" terms in Feynman diagrams.   In operator formalism, these constraints appear as the Ward-Takahashi identity at operator level
\begin{equation}
  \label{wardtak}
\int \mathcal{D}[A,\psi] \exp\left[ S[A,\psi\rightarrow A,\psi+\partial_\mu j^\mu \right] = \ave{\hat{k}_\mu \hat{\mathcal{M}}^{\mu}}=0
\end{equation}
where $\hat{\mathcal{M}}^\mu$ is an operator that creates an outgoing  field of polarization in the $\mu$ direction.

We now realize that the integrand over \eqref{zuberrho} is done over $\mathcal{D}\phi$ but the exponent occurs over $T_{\mu \nu}$.    The reparametrization symmetry we advocate is exactly over ensembles of configurations over $\phi$ that leave $T_{\mu \nu}$ unaltered.  In the coarse-grained \eqref{zuberrho} this is done by a simultaneus redefinition of $d\Sigma_\mu$ and $\beta_\mu$ via transformations of the type
\begin{equation}
  \label{msigmadef}
  d\Sigma_\mu \rightarrow M_{\mu \nu}d\Sigma^\nu \eqcomma \beta_\mu \rightarrow M_{\mu \nu} \beta^\nu \eqcomma M_{\alpha \beta} \frac{\delta \lnz}{\delta g_{\alpha \nu}}=0
\end{equation}
where the last equation is a direct analogy of the Gauge constraint equations of the type \eqref{ward}.
Note that temperature (the normalization of $\beta_\mu$), $u_\nu$ and of course $d\Sigma_\mu$ are shifted separately.\color{black}  Note also that \eqref{ward} refers to the observable $T_{\mu \nu}$, while the transversality condition on the unobservable \eqref{trans} is generally broken by the reparametrizations;  As a rough analogy to the fact that in gauge theory virtual bosons can have unphysical polarizations. \color{black}

Since these transformations are generally non-inertial, the constraint on energy momentum conservation is in the form
\begin{equation}
  \partial_\mu \ave{T^{\mu \nu}}+\Gamma_{\sigma \rho}^\mu\ave{T^{\sigma \rho}}=0
  \label{conserv}
  \end{equation}
which is of course the origin of \eqref{ward}.   Note however that the first term is generated by the second term of \eqref{zuberrho2} while the second term is generated by the first.    It is therefore clear that transformations of the type represent the contact terms of \eqref{ward}.     From the point of view of the transformed coordinate system $\beta'$ one can see that
	\begin{equation}
		\label{constraint}
		\int_{t_0}^{t} \!\!\! \di^4 x'  \int_{t_0}^{t'} \!\!\! \di \theta \; \ave{
			\left[ \left( M_{\alpha \zeta} \wT^{\zeta \kappa}(x)\right),\wT^{\mu\nu}(\theta,{\bf x}')\right] }_{\beta(x)} \partial^\alpha  \beta_\kappa'(x')  =0\,,
	\end{equation}
which can be used to find possible $M_{\alpha \beta}$.
	But it is obvious that $\Pi_{\mu \nu}$ and $\beta_\mu$ do change.  Note that if one keeps just the averages dynamics will generally change since $M_{\alpha \beta}$ is defined, via
\ref{avconst}, in terms of {\em average} $\ave{T_{\mu \nu}}$.  {\em
Ensembles} such as $\beta^\mu T_{\mu \nu}(\phi)$ for a fluctuating
configuration of $\phi$ in \eqref{zuberrho} generally transform
non-trivially.

This explains why the dynamics within \cite{bdnk} and \cite{IsraelStewart} is different beyond leading order\color{black} and why field redefinitions generally cascade into higher order terms in gradient (as they do in \cite{yarom}).  A consistent treatment including fluctuations and dissipation (analogous to the Dyson-Schwinger equations in field theory) would recover a dynamics invariant under parametrizations.   As discussed earlier such a dynamics would depend on the contact terms in \eqref{ward} and would mix dissipative and non-dissipative parts of \eqref{zuberrho2}.

	The above formula \color{black} allows us to translate Kubo formulae from IS throughout the flow.  The trace of this formula gives the bulk viscosity while the traceless part is the shear tensor.

	and \eqref{ward}, \eqref{constraint} provide the equations necessary to translate \eqref{eq:transportkubo} from a Landau frame to a local frame defined in term $M_{\alpha \beta}(x)$.

        \color{black}
        In the limit of small coarse-graining of \cite{crooks}, we can implement the condition of \eqref{crooks} by absorbing the non-equilibrium term in the partition function
        \begin{equation}
\label{eomz}
          \mathcal{Z}(\tau,\Sigma,\beta)= \int \mathcal{D}\phi \exp 
\left[ -\int d \Sigma_\mu(\tau) \left(  \beta_\nu \hat{T}_0^{\mu
\nu} + \beta^\mu \hat{\Pi}^{\alpha \beta}
    D_\alpha \beta_\beta \right) \right]
        \end{equation}
        where $D_\alpha$ is the covariant derivative and all scalars are computed with the co-moving metric.
        
        it is of course strange to see a partition function depending on time, but we must remember that this is a dissipative system based on \cite{crooks} the separation between the Markovian ``decoherence'' scale and the macroscopic scale.  It is therefore appropriate to describe it by a time-dependent partition function, provided covariance between different choices of $d^3 \Sigma_\mu$ is respected.                If one defines a metric $g_{\mu \nu}=\partial \Sigma_\mu/\partial \Sigma^\nu$ and applies \eqref{conserv}, it is clear that $M_{\alpha \beta}$ represents exactly these transformations. \color{black}

	{ If one models the thermal fluctuation \cite{kadanoff} as an external perturbation \[\ H (t) =  - \int^{\infty}_{- \infty} dt \ \bigg[ \frac{\delta T}{T} (t) \epsilon (t) + v_i (t) \pi (t) \bigg]  e^{ \epsilon t}\]. In the limit $\vec{k} \rightarrow 0$  the correlation function of irreversible processes has the form 
		\begin{equation}
			\label{gepsilon}
			\begin{aligned}
				G^{\epsilon \epsilon} (\omega, 0) &= \langle \epsilon \rangle  - \Re \{ |\frac{\delta T}{T} (x^\mu)| \xi(\omega) e^{i\omega t + \tau t}  \} \\ 
				G{\pi^j \pi^i} (\omega, 0) &= - \delta_{ij} \langle \epsilon \rangle + \Re \{ |v_i (x^\mu)| v_s (\omega) e^{i\omega t + \tau t}  \}
		\end{aligned}\end{equation}
		where $\Re$ is the real part and $v_s$ is the sound velocity. In the limit $\omega \rightarrow 0$
		\begin{equation}
			\label{gpiw}
			\begin{aligned}
				G^{\pi^j \epsilon} (\omega, 0) &= \langle \epsilon \rangle  - \Re \{ |v_i (x^\mu)| \xi (\omega) e^{i\omega t + \tau t}  \} + \chi_{\pi \pi} (\vec{k}) \\ 
				G{\pi^j \pi^i} (0, \vec{k}) &= - ( \delta_{ij} \delta_{lm} + \delta_{il} \delta_{jm} + \delta_{im} \delta_{nj} )  \langle P \rangle  + \Re \{ |v_i (x^\mu)| \nu (\omega) e^{i\omega t + \tau t}  \} +\chi_{ijlm} (\vec{k})
		\end{aligned}\end{equation}
		where $\chi$ are function at least  $\mathcal{O}(\vec{k^2})$ in the long wavelength limit. Taking the canonical ensemble to average over the energy and pressure operators.  Note that the $\delta_{ij}\delta_{jm}$ terms are of the form of \eqref{jproj}}   Such an effective expansion is described in more detail in the next section.
	
	\subsection{Reparametrization, transport theory and Gibbsian entropy}\label{reptransport}
	It is clear that \eqref{flowstep} and \eqref{bulktrans} mix microscopic fluctuations and hydrodynamic correlations 
	In particular, as is clear from \cite{montenegro,crooks}, the hydrodynamic limit works when the hierarchy
	\begin{equation}
		\label{hyerarchy}
		\underbrace{T_0^{-1}}_{\sim s^{-1/3} \sim T^2/(c_V)} \ll \underbrace{l_{\rm mfp}}_{\eta/(Ts)} \ll \left( \partial u_{\mu} \right)^{-1}\,.
	\end{equation}
	Here $l_{\rm mfp}$ is the mean free path length.
	The first scale regulates thermodynamic fluctuations \color{black} and appears, in ideal fluctuating hydrodynamics, as a ``Planck constant attached to an energy scale'' in a functional integral defined via Lagrangian coordinates $\phi_I$ \cite{ryblewski}
\[\   \lnz \sim \int \mathcal{D} \phi_I \exp T_0^4 S[...]  \]
, and the second dissipation (where it appears as a gradient term).   \color{black} The extra scale represented by the first inequality is usually ``swept under the rug'', since in both the Boltzmann equation and gauge/gravity duality it is manifestly zero by, respectively, molecular chaos and the planar limit.   It's existence however is manifest by the \color{black}contact terms in \eqref{ward}.  As   \eqref{eq:transportkubo} is defined in the infrared limit (where fluctuations are negligible) while \eqref{eq:correlator} is a complex object which in principle contains arbitrary structures in $k,w$.   Taking the infrared limit and assuming the baclground hydrostatic makes all fluctuation terms, and any effect of the contact terms in \eqref{ward} disappear, and in the planar/molecular chaos limit one can safely neglect them for all values of the gradient.

However, when viscosity and the fluctuation scale are comparable and the background is not hydrostatic contact terms in \eqref{ward}, representing autocorrelations within a non-static background are non-trivial.   One way to account for them is to change
Correlators such as \eqref{eq:correlator} to a metric deviating from $\eta_{\mu \nu}$.  The non-inertial forces \color{black} can be interpreted as a fluctuation or as a correlation depending on the time-direction of the foliation frame (which chooses the real part, interpreted as a fluctuation vs. the imaginary part, interpreted as a dissipation).   The distinct ways one can arrive at the definition of the relaxation time, one explicitly dissipative \cite{jeon,dirkis} and the other thermodynamical \cite{hislim,lindblom} are an illustration of this ambiguity.
        Because physics is invariant under Lorentz symmetry and, when close to local equilibrium it is invariant under foliation choice, when the first inequality fails but the second holds one should therefore rotate across the fluctuation-dissipation axis.
        
        This is exactly what
        Equation~(\ref{constraint}) does: it {\em rotates} in an axis $P^\mu$ defined by the first hierarchy, in terms of unit vectors $\hat{n}_{1,2}$
	\begin{equation}
		\label{hyerarchym}
		P^\mu(x) = \left(\begin{array}{c} T_0^{-1} \hat{n}_1 \\
			l_{mfp} \hat{n}_2\end{array} \right)
		\rightarrow P^\mu(x) +M_{\nu}^\mu(x)P^\nu(x)\,,
	\end{equation}
The usual gradient expansion has $P^\mu$ pointing along the $l_{mfp}$ direction, so all anisotropy is due to dissipative effects.  If $T_0$ is comparable to $l_{mfp}$, this is no longer true, but it becomes possible to
rotate $P_\mu$	resumming thermal and dissipative modes so as to leave the fluid invariant.
Thus, thermal fluctuations (forming at a scale $T_0$) and sound waves (dissipating at a scale $l_{mfp}$) become "Gauge dependent", as expected since if $T_0^{-1}$ is comparable to $l_{mfp}$ they are indistinguishable.\color{black}.

	Understanding the effective expansion around this resummation requires
	an effective theory of hydrodynamic fluctuations~\cite{stick,kovfluct,hnatic} as well as a microscopic description.  In particular, a Boltzmann equation description (where the first scale in the hierarchy is zero) is likely to fail when turbulence causes the two scales to be significantly mixed.  This is obvious from the fact that the H-theorem, inherent in the Boltzmann dynamics, is not always respected within anisotropic hydrodynamics and is a manifestation of the fact that the Boltzmann and Gibbs entropy could be significantly different in non-equilibrium systems \cite{jaynes}.     The Gibbs entropy  indeed provides a physical picture of \eqref{flowstep} and \eqref{bulktrans}, and their relationship to the tensions discussed after \eqref{eq:transportkubo}.  The entropy current counts the microstates which are compatible with a given macroscopic information, but not {\em all} macrostates compatible with a given value of $\Pi_{\mu \nu}$, but {\em only those} compatible with the way the system is perturbed, which are a fraction of microstates where $\Pi_{\mu \nu}$ has this given value. Equation~(\ref{avconst}) provides a way to directly probe how the number of microstates responds to a given perturbation.
	On the other hand, since $\Pi_{\mu \nu}$ is not observable itself, it should not be considered a ``microstate'' in the Gibbsian sense.  Instead, \eqref{flowstep} and \eqref{bulktrans}, together with \eqref{avconst} define the non-equilibrium phase space area compatible with a given $\Pi_{\mu \nu}$ field. There are generally multiple ways to parametrize this phase space, and this is precisely the physical meaning of frame choice.
	However, the non-equivalence of Gibbsian and Boltzmannian entropy means that the transport coefficients cannot be understood by the Boltzmann equation alone, even when thermal contributions are resummed~\cite{arnold}.
	This is because the interaction corrections to microscopic fluctuations, at all scales, are neglected in the molecular chaos assumption.
 
\color{black}To clarify the above statement, we note that a recent truncation of the Boltzmann equation proposed in \cite{denicol} has been used to derive anisotropic hydrodynamics in \cite{kovtunani} (a renormalization group treatment still seems to require isotropy \cite{kunihiro2}).    However, as explained in the appendix \ref{apptoy}, one loses control of the Boltzmannian entropy definition in this truncation, as consistent with the fact that anisotropic hydrodynamics occasionally and temporarily breaks the H-theorem \cite{GavaCausality2021} while the Boltzmann equation obeys it exactly.   Since entropy content drives fluctuations, and the fluctuation-dissipation theorem controls dissipative terms within hydrodynamics, this illustrates that a consistent microscopic completion of hydrodynamics needs both.
	
	\color{black} 
	In the semi-classical limit, microscopic correlations in the infrared are best modeled via a Vlasov term, of which the Boltzmann term can be regarded a UV completion \cite{villani} (random scattering UV-completes filamentation, where more and more local dynamics becomes more and more unstable).    More concretely, the two term represents  the interplay of the thermal ($\sim T$), Debye ($\sim g T$), and transport ($\sim g^2 T$) perturbative scales studied in~\cite{arnold}.  In perturbation theory (where the coupling constant $g \ll 1$) the smallest scale regulates microscopic fluctuations, while the largest scale regulates the mean free path (and ultimately the Knudsen number).  The Debye scale regulates soft scatterings, which are problematic because the quantum duration of the scatterings is not necessarily microscopic.
        However, they can usually be absorbed into a Vlasov-type mean field\cite{crismanuel}.
        
        As shown in~\cite{arnold}, when the cumulative effect of soft scatterings is much smaller than the hard scatterings, close to the hydrostatic limit, they can be ressumed into the parameters of an effective Boltzmann equation.  This, however, introduces ambiguity when the Debye screening length and the viscosity are not so well-separated.  While many soft scatterings do not add up to a hard (momentum exchange $\sim T$) scattering affecting transport, they might well add to a change in momentum comparable to a boost by $u_\mu$.  In this limit, such a microscopic-macroscopic correlation is of the order of the Knudsen number, and hence the effective Boltzmann description breaks down, and needs to be augmented by a Vlasov term.

        When the number of particles is small and thermal fluctuations become non-negligible, ensemble average breaks down  and the microscopic distribution function $f(x,p)$ should be generalized to a {\em functional } $\mathcal{F}(f(x,p))$ of all possible distribution functions compatible with a given event distribution, which in the limit where the ensemble average is a good approximation tends to $\mathcal{F}(f(x,p)) \rightarrow \delta(f-f(x,p))$ .   This is the semiclassical limit of the Wigner functional discussed in \cite{wignerf} if coherence between different volume cells is negligible.   

        In this regime one could write the correction to the Boltzmann equation as\footnote{Note that the fact that the Vlasov term is on the RHS breaks Liouville`s theorem.   Since the LHS represents a function, and the RHS a functional, this is to be expected, and can be understood by the fact that the phase space area within this limit includes ``all possible distribution functions`` compatible with a sample away from the ensemble average.  The collision term as a UV completion, argued for in \cite{villani} can also be understood this way.  }
\begin{equation}
\label{vlasovfunc}
\frac{p^\mu}{\Lambda} \frac{\partial}{\partial x^\mu} f(x,p)=\ave{
  \hat{C}[f_1,f_2)]
-g \hat{V}\left(f_1,f_2\right) }
\end{equation}
	where $\hat{C}$ is the Boltzmann collision operator and $\hat{V}$ the Vlasov collision operator, both based on te Wigner functional \cite{wignerf}\footnote{their exact form is left to a forthcoming work} and the $\ave{...}$ is performed over all possible $f_{1,2}$.   
We purposefully put the Vlasov collision operator on the RHS of the equation as it makes the origin of the gauge-like ambiguities clear.

As can be seen when the Boltzmann term only appears, one can expect the relaxation of a function to a functional would entail a small correction if $\mathcal{F}(f(x,p)) \simeq \delta(f-f(x,p))$ .   The {\em difference} between a Boltzmann and Vlasov terms however introductes an additional ambiguity, since Boltzmann is sensitive to configuration space deformations and Vlasov to momentum space deformations.

There will therefore generally be an infinity of possible deformations,
\begin{equation}
  \label{collinv}
f(x,p) \rightarrow f'(x,p) \eqcomma \underbrace{\hat{C}\left(f(x,p),f'(x,p)\right)}_{lim_{f \rightarrow f'} \sim \partial f/\partial x} = \underbrace{ \hat{V}\left(f(x,p),f'(x,p) \right)}_{lim_{f \rightarrow f'} \sim \partial f/\partial p}  \end{equation}
{\em both} in position and momentum space, that leave the RHS of \eqref{vlasovfunc} unchanged.   In particular, in the Boltzmann equation case the RHS of \eqref{vlasovfunc} would only vanish in the opposite limits of free-streaming and ideal fluid dynamics.   But for  \eqref{vlasovfunc}, when both the Boltzmann and Vlasov term are included, the RHS vanishes for a whole class of redundant configurations, that does not vanish when the functional narrows around an average function.

Furthermore, assuming the system is close enough to equilibrium, so that
    \begin{equation}
      \label{vlaszub}
      \left\{
\begin{array}{c}
  f(x,p)\\
  f'(x,p)
\end{array}
\right\} \sim \exp\left [-\left\{
\begin{array}{c}
  \beta_\mu(x,t)\\
  \beta'_\mu(x,t)
\end{array} \right\}
p^\mu \right] \eqcomma \lim_{f \rightarrow f'} \left\{ \begin{array}{c}  \hat{C}[f,f']\\
  \hat{V}[f,f']
\end{array} \right\} \sim \left\{
\begin{array}{c}
\ave{\partial \beta} \\
\ave{\beta^2} \end{array}
\right\}
\end{equation}
  and comparing to \eqref{zuberrho} it is clear that changes in $f$ can be reduced to fluctuations in $\beta_\mu$ and changes in foliation.  Since these are captured by \eqref{flowstep} and \eqref{bulktrans}, we can identify these ambiguities with the gauge-like ambiguities examined in the previous section.\color{black}

Hence, an anisotropic $u_\mu$, for perturbative systems close to equilibrium, is a way to absorb the breakdown of molecular chaos due to Debye and transport scales being not so well separated.    A similar picture emerges in non-relativistic fluid dynamics~\cite{trachenko} close to minima of viscosity, with the scale associated with Frenkel's phonon states replacing Debye screening lengths. 
	\subsection{Reparametrization and the effective theory of hydrodynamics \label{justeft}}
	From a ``Wilsonian renormalization'' point of view, hydrodynamics is an effective theory and the Gibbsian entropy is related to the measure of the effect of the choice of coarse-graining scale on the parameters of the effective theory, which for hydrodynamics appear as linear response \eqref{constraint}\cite{kadanoff}.

        \color{black}
  In case of transport, one can think of the Boltzmann term as an ultraviolet completion of the Vlasov terms (where more an more unstable dynamics at finer and finer scales is continued into random scattering for infinitesmical scale).  In this case, each term on the right hand side of  \eqref{vlasovfunc} is scale dependent, but the difference is not.   Thus, the redundances discussed in this work survive coarse-graining.\color{black}
        
	However, the coarse-graining scale of fluctuation and dissipation processes are generally different. One is the ``micro'' scale and the other is the mean free path scale. This means that a continuum of equivalent theories is possible where fluctuation and dissipation ``conspire'' to produce the same macroscopic dynamics, in a way that is hidden in how the two microscopic scales interact.  
	The different frames of hydrodynamics is a manifestation of this, where the analogy between renormalization group flow and hydrodynamics~\cite{rghydro} is used to evolve one coarse-graining scale in the space of theories, and the other as a time-scale.
	Unlike gauge theory, hydrodynamics is neither unitary nor renormalizable;  Therefore, one cannot hope gauge dependence will cancel order-by-order when diagrams are summed (a ``tree-level'' example is Fig. \ref{feyndiag}).   Instead, one adjusts effective field theory parameters, i.e. Equation of state (EoS), viscosity, and so on, so they give the same order-by-order \eqref{constraint}. The criterion of stability in selecting between the theories can be justified as a way to choose a critical surface ($h_\epsilon$ in~\cite{spohn}) with the perturbation given by thermal perturbations. At least for finite dimensions, one can prove that as long as fluctuations are small such a choice is always possible order-by-order.     
	
	In other words, to convert the effective theory coefficients across this space of theories the interaction between sound waves, microscopic thermodynamic fluctuations, and dissipation need to be included within each coarse-graining scale.  If the effect of one is forgotten, the transport coefficients will be incorrect, and the separation of scales in \eqref{hyerarchy} by itself is not enough to ensure any effective theory will be equally insensitive to microscopic parameters. Equations (\ref{zuberrho}-\ref{noneqent}) make it clear that this dependence comes from the non-commutativity of the choice of $d^3\Sigma_\mu$ and the gradient expansion in $\beta_\mu$.  The constraint \eqref{constraint} allows us to change both in a way that the effective theory in the infrared is the same.
	
	This places some limits on the applicability of such generalized effective theories.  For instance, in the large $N_c$ limit (or, generically, in the large degeneracy limit), the rotation in \eqref{hyerarchym} would inevitably be $N_c$ suppressed. In the next section, we demonstrate this explicitly in the case of bulk viscosity. This is because microscopic degrees of freedom go as $g\sim N_c^2$, while a gas of sound wave and vortex excitations inevitably has degeneracy of unity. The first scale, therefore, collapses to ``zero'' and any rotation between the first two scales destroys the applicability of classical gauge gravity duals.
	
	Ultimately, while the expansion in the second hierarchy of \eqref{hyerarchy} generates dissipative terms, the first hierarchy generates stochastic terms~\cite{crooks}. Mixing the two would require functional integral techniques, such as \eqref{zuberrho} and the methods of Sec.~\ref{functional}. The fact that entropy in the first-order approach sometimes oscillates \cite{gavassino}, as expected from microscopic fluctuations \cite{crooks} is a manifestation of this.
	
	Equation (\ref{zuberrho}) makes it clear that one can absorb $M_{\alpha \beta}$ in a reparametrization of $d\Sigma_\mu$ and the associated boundary term $\int d\Omega$. These are arbitrary, not necessarily subject to the constraint of $\Sigma_\mu \propto u_\mu$ (this would correspond to the co-moving frame in the ideal hydrodynamic limit).
	A generic decomposition respecting Equations (\ref{flowstep}) and (\ref{avconst}) would automatically implement
	\begin{equation}
		\label{intconst}
		\int \mathcal{D}\phi \rightarrow \int \mathcal{D} \alpha_{I=1,2,3} \mathcal{D} \beta_{I=1,2,3} \  \mathcal{D}\left[ \mathcal{A},e,p,u_\mu,\Pi_{\mu \nu}\right]  \delta\left( M_{\alpha \beta}\left[\mathcal{A},\alpha_I,\beta_I  \right] T^{\alpha \mu} \right)
	\end{equation}
	where $\alpha,\beta,\mathcal{A}$ are auxiliary ghost fields.   Note that this is the largest number of degrees of freedom possible consistent with the condition of \eqref{zuberrho} that the exponent in the density matrix only depends on $T_{\mu \nu}$, since the argument of the $\delta(...)$ leaves it invariant.
	
	A BDNK type partition function would correspond to a choice of $d\Sigma_{\mu}$ for which 
	\[
	\Pi_{\mu \nu}  \nabla^\mu \beta^\nu=0\,.
	\]
	The second integral would therefore become a purely ``constraint'' one, similar to a ghost constraint integral in Gauge theory.
	\subsection{Reparametrization and causality \label{repcause}}
	Taking these symmetries into account at the ensemble level is essential for a discussion of causality. One notes that Equations (\ref{eom}) and (\ref{constraint}) refer to {\em time-ordered commutators} rather than simple fluctuations. As we know from quantum field theory, it is the former, rather than the latter, that need to be causal. Precisely and in analogy to quantum field theory
	\begin{equation}
		\label{causalcond}
		\ave{\tcomm{ \delta T_{\mu \nu}(x)\delta T_{\mu \nu}(x')}}\equiv \ave{T_{\mu \nu}(x_1,t_1),T_{\mu \nu}(x_2,t_2)}- \ave{T_{\mu \nu}(x_1,t_2),T_{\mu \nu}(x_2,t_1)}=0
	\end{equation}
	\[\   \Leftarrow |x_1-x_2|-|t_1-t_2|\geq 0 \]
	Here $\tcomm{...}$ denotes the time-ordered commutator brackets.

	Because of thermal fluctuations 
	$\ave{T_{\mu \nu}(x_1,t_1),T_{\mu \nu}(x_2,t_2)}$ could have non-zero space-like correlations which a deterministic theory would interpret as non-causal.   The situation is further complicated if one considers that $p,\,e,\,u_\mu$, and $\Pi_{\mu \nu}$ are {\em not} observables.
	If one considers the before-mentioned hydro variables as precise quantities evolving deterministically, Landau matching is unique and lack of
	causality in the propagation of any of the hydro variables means that the theory is not causal.
	However, $p,\,e,\,u_\mu$, and $\Pi_{\mu \nu}$ under \eqref{zuberrho} are  ensembles obtained by decomposing an ensemble of $T_{\mu \nu}$ under some choice according to \eqref{liegroup}.
	Hence, superluminal propagation of an object like $\ave{\delta u_\mu(x) \delta \Pi_{\mu \nu}(x')}$ or even $\ave{\tcomm{\delta u_\mu(x) \delta \Pi_{\mu \nu}(x')}}$  is {\em not} acausal provided that $\ave{\tcomm{ \delta T_{\mu \nu}(x)\delta T_{\mu \nu}(x')}}$ in \eqref{causalcond} is causal and \eqref{ward} is respected.
	Hence, an integration of the ensemble of \eqref{zuberrho} under \eqref{intconst} is necessary to understand causality of this system. Non-causal correlations of $\alpha_I$ and $\beta_I$, for instance, could be balanced by non-causal correlations of $\Pi_{\mu \nu}$ and $u_\mu$ to yield a causal propagation of $\ave{\tcomm{ \delta T_{\mu \nu}(x)\delta T_{\mu \nu}(x')}}$, in analogy with the effective theory discussed in~\cite{glorioso}.  Equation (\ref{intconst}) together with a \eqref{zuberrho} partition function makes sure that any fluctuations that make $T_{\mu \nu}$ invariant was out in ensemble averaging. 
	
	Physically, this situation is a reflection of the fact that it is impossible to distinguish ``event by event'' the contribution to  a thermodynamic fluctuation (which need not be ``causal'') from a deterministic evolution (which is).  Acausality becomes important when the fluctuation probability becomes negligible.
	As noted a long time ago~\cite{ruderman}, the onset of acausality could coincide with a regime where the contribution to microscopic fluctuations is non-negligible.  In this case, it is this contribution that must be properly modeled to get causality back.
	
	We can use the results of the deterministic calculations in~\cite{spal} to calculate the theory to zeroth order, when the functional integral reduces to an ordinary integral over $d U_{\mu \nu}$
	\begin{eqnarray}
		\label{black}
		T^{\mu \nu}(x,t) &=& \int^t_{-\infty} dt' \int  d \mathcal{A} \prod_{I=1}^3 d\alpha_I d\beta_{I} \prod_{\kappa=1}^4 \delta \left( U_{\alpha \beta} \left(\mathcal{A},\alpha_I,\beta_I \right)T^{\alpha \kappa}(x,t')
		\right) \times \nonumber \\ &&
		\sum_{I=1}^6 \left( \mathcal{P}^{-1}_I \right)^{\mu \nu} \int d^3 k e^{ikx}   \tilde{T}_{\alpha \beta}(k,t') G_I(k) \mathcal{P}_I^{ \alpha \beta}
	\end{eqnarray}
	Here, $G_I(x,x')$  are the propagators of the sound and diffusion modes (given by the dispersion 10-16 of~\cite{spal})
	\begin{equation}
		\label{propagator}
		G_I(k) =  \frac{1}{\Gamma_I (k)} \eqcomma \Gamma_I(k) \sim (a_1(T,\tau)+ib_1(T,\tau)) k+(c(T,\tau)+id(T,\tau))k^2\,,
	\end{equation}
	In \eqref{black}, the sum $I$ projects over Eqs. (\ref{jproj}) and (\ref{jproj}), $U_{\mu \nu}$ is the transformation generated by $\mathcal{A},\alpha_I,\beta_I$\footnote{Since this is linearized and defined in the co-moving frame $\Lambda_{\alpha \beta}$ are unity so it is given by \eqref{element2}}, and $\mathcal{P}_I^{\mu \nu}\left(\mathcal{A},\alpha_I,\beta_I \right)$ projects the $T_{\mu \nu}$ tensor onto the $I$th component. The beforementioned projector reads 
	\begin{equation}
		\label{pdef}
		\mathcal{P}_I^{\mu \nu} = \left( \exp\left[ \mathcal{A}g^{\mu \nu} + \sum_J \alpha_I J^{\mu \nu}+\beta_J K^{\mu \nu} \right] \right)_{ij(I)} \eqcomma ij(I) = \begin{array}{cc}
			ij(I) & I \\ \hline
			00 & 1\\
			0 \hat{k}_\parallel & 2\\
			0 \hat{k}_{1\perp} & 3\\
			0 \hat{k}_{2\perp} & 4\\
			\hat{k}_\parallel \hat{k}_{1\perp} & 5\\
			\hat{k}_{1\perp} \hat{k}_{2 \perp} & 6\\
		\end{array}
	\end{equation}
	and $\hat{k}_{...}$ are unit vectors parallel and perpendicular to the sound wave.
	$\left(\mathcal{P}_I^{-1}\right)^{\mu \nu}$ builds the tensor back. In the small $k$ limit these equations converge to \eqref{gepsilon} and \eqref{gpiw}.
	
	The integrand goes over every Gauge configuration compatible with the initial $T_{\mu \nu}$. Linearized causality analysis should therefore be done on \eqref{black} rather than on the dispersion relations of~\cite{spal}.  Note that to zeroth order, the $\delta(...)$ function factors into the initial condition. Backreaction will alter this and are expected to be significant even in the linearized limit since in any case this limit gets admixtures from thermal fluctuations as per \eqref{hyerarchym}.

	Physically, the above discussion is a reflection of the ambiguity between $T_0^{\mu \nu}$ and $\Pi^{\mu \nu}$ once thermal fluctuations are taken into account. In linearized hydrodynamics, once "the background" around which we are expanding is not known precisely (i.e. unless the fluctuation scale and the dissipation scale are well separated) the oscillating transient modes in~\cite{janik} are in principle indistinguishable from propagating sound waves. In fact, by changing $d \Sigma_\mu$ in the Zubarev picture from moving along the sound wave to a perpendicular direction, the transient mode in $\Pi_{\mu \nu}$ will become part of the oscillating sound wave.
	Consequently, the oscillating transport coefficient can be interpreted as a relaxation time or a fluctuation-driven input of the equation of state. The necessity of including all fluctuation terms is apparent when one sees the pole structure of the dispersion relations discussed in~\cite{janik}.  The pure imaginary locations of poles of Boltzmannian transport theory are a reflection of the H-theorem, which is a consequence of the fact that only 1-particle distributions, disregarding fluctuations and correlations, are included.  AdS/CFT changes this in a very particular way (fluctuations are still suppressed, but planar correlations are not), and as can be seen in~\cite{janik}, this is enough to shift the pole structure to the real line.  Generally, \eqref{constraint} means the dispersion relation must be invariant under shifts which leave the total $T_{\mu \nu}$ invariant, whose analytical continuation generally mixes real and imaginary directions. This is an additional indication that relying on an arbitrary separation between $T_0^{\mu \nu}$ and $\Pi^{\mu \nu}$ could lead to physically incorrect insights.
	
	\section{The case of pure bulk viscosity \label{bulkvisc}}
	For the case of pure bulk viscosity, we can build a BDNK model as a field theory for the fields 
	\begin{equation}
		(u^\mu,T_B),
	\end{equation}
	decomposing the stress-energy tensor as follows:
	\begin{equation}
		\label{stressdeco}
		T^{\mu \nu} = (e_B + \mathcal{A})u^\mu u^\nu + p_B(u^\mu u^\nu - g^{\mu \nu}),
	\end{equation}
	and imposing the constitutive relations
	\begin{equation}\label{constitutiveBDNK}
		p_B =p_B(T_B)  \quad \quad dp_B = s_B dT_B \quad \quad e_B = s_B T_B -p_B  \quad \quad \mathcal{A}= \chi_1 \frac{\dot{e}_B}{e_B +p_B} + \chi_2 \nabla_\mu u^\mu.
	\end{equation}

	The IS theory, on the other hand, builds on three fields, 
	\begin{equation}
		(u^\mu,T_I,\Pi),
	\end{equation}
	and assumes the following decomposition:
	\begin{equation}
		\label{isdeco}
		T^{\mu \nu} = e_I u^\mu u^\nu + (p_I+\Pi)( u^\mu u^\nu -g^{\mu \nu}),
	\end{equation}
	with constitutive relations
	\begin{equation}\label{ConstitutiveIS}
		p_I =p_I(T_I)  \quad \quad dp_I = s_I dT_I \quad \quad e_I = s_I T_I -p_I 
	\end{equation}
	and an additional Maxwell-Cattaneo-type field equation \cite{JouBulk}
	\begin{equation}\label{IsraelRelax}
		\tau_\Pi \dot{\Pi} + \Pi = -\zeta \nabla_\mu u^\mu.
	\end{equation}
	The equilibrium equation of state $p_I=p_I(T_I)$ is assumed to be the same as the BDNK one, $p_B =p_B(T_B)$.

	\subsection{Matching the two theories\label{match}}
	
	Since the stress-energy tensor must be the same for the two theories, we immediately have the identifications
	\begin{equation}
          \label{identifications}
		e_B +\mathcal{A} = e_I  \quad \quad \quad p_B = p_I +\Pi.
	\end{equation}
	Truncating these equations to the first order in deviations from equilibrium, i.e. in $T_B-T_I$, we obtain the bridge formula
	\begin{equation}\label{BridgeFormulaAA}
		T_B - T_I = \frac{\Pi}{s_I} = -\frac{\mathcal{A}}{c_v^I}  \quad \quad \quad \text{with} \quad \quad \quad c_v^I := \frac{de_I}{dT_I}.
	\end{equation}
	We notice that the correction is inversely proportional to the degeneracy which enters $C_V$.  This is a confirmation of the hierarchy \eqref{hyerarchym}, where it is shown that these transformations are a case of microscopic and macroscopic fluctuations being mixed.  For instance, in AdS/CFT $c_V$ is parametrically large so any difference betweeen $T_{B,I}$ is bound to be negligible inasmuch as AdS/CFT is a good approximation.
        	
	Combining this formula with \eqref{constitutiveBDNK} and truncating at the leading order we find
	\begin{equation}
		- \frac{c_v^I}{s_I} \, \Pi = \chi_1 \frac{\dot{e}_I + c_v^I \dot{\Pi}/s_I }{e_I+p_I} + \chi_2 \nabla_\mu u^\mu.
	\end{equation}
	Comparing with \eqref{IsraelRelax} we are able to write $\chi_1$ and $\chi_2$ is terms of Israel-Stewart coefficients:
	\begin{equation}\label{matchingCoefficients}
		\chi_1 = (e_I +p_I) \tau_\Pi \quad \quad \chi_2 - \chi_1 = \frac{c_v^I}{s_I} \zeta. 
	\end{equation}
	Indeed, the rest-frame stability conditions for this BDNK model are
	\begin{equation}
		\chi_2 > \chi_1 >0,
	\end{equation}
	which correspond to the rest-frame Israel-Stewart stability conditions
	\begin{equation}
		\tau_\Pi >0  \quad \quad \zeta >0.
	\end{equation}
	The bridge formula, \eqref{BridgeFormulaAA}, can also be used to compare the entropy currents of the two theories:
	\begin{equation}
		\label{entropycomp}
		s_B^\mu = \bigg( s_B + \frac{\mathcal{A}}{T_B}  \bigg)u^\mu  \quad \quad \quad s_I^\mu = s_I u^\mu.
	\end{equation}
        \color{black}
        As subsection \ref{reptransport} makes clear, the {\em Gibbsian} entropy should be invariant under these reparametrizations. Equation \ref{entropycomp} should therefore be seen as a recipe to relate Gibbsian and Boltzmannian entropies of the two pictures.   Indeed, if one looks at the entropy density one sees that
	\begin{equation}
          \label{entropydensitycomp}
		s_B-s_I = \frac{c_v^I}{T_I} (T_B-T_I)= -\frac{\mathcal{A}}{T_I}= -\frac{\mathcal{A}}{T_B},
	\end{equation}
        hence the $s_{B,I}$ redefinition of \eqref{entropydensitycomp} exactly compensates the shift in $p,e$ to make the two terms of \eqref{entropycomp} the same.    Physically, these redefinitions represent re-labeling the backreaction to microscopic fluctuations as ``macro'' vs ``microstates'', with the fluctuations themselves being re-labeled as ``heat currents'' vs ``collective modes''.   In the next subsection we will make this connection more explicit\color{black}
	\subsection{Spectral analysis\label{spectral}}
	
	
	Let us linearise (for simplicity in 1+1 dimensions) the equations of both theories around a static homogeneous equilibrium state. Then we compare the spectra of the excitations assuming a space-time dependence of the perturbations of the form $\exp(ikx-i\omega t)$. For BDNK, we have
	\begin{equation}
		\delta T^{00} = \delta e_B + \delta \mathcal{A}  \quad \quad \quad  \delta T^{01} =(e_B  + p_B) \delta u^1  \quad \quad \quad  \delta T^{11} = \delta p_B \, ,
	\end{equation}
	so that $\nabla_\mu T^{\mu \nu}=0$ becomes
	\begin{equation}
		-i\omega (\delta e_B + \delta \mathcal{A}) + ik (e_B  + p_B) \delta u^1 =0  \quad \quad \quad -i\omega (e_B  + p_B) \delta u^1 + ik \delta p_B =0.
	\end{equation}
	Combining these field equations with the constitutive relation
	\begin{equation}
		\delta \mathcal{A}= -i \omega \, \chi_1 \frac{\delta e_B}{e_B +p_B} +ik \, \chi_2 \delta u^1
	\end{equation}
	we obtain the condition
	\begin{equation}\label{spectrumBDNK}
		\frac{\chi_1}{e_B+p_B} \, \omega^3 + i \omega^2 - \frac{\chi_2 s_B}{(e_B+p_B)c_v^B} \, k^2 \omega - i\frac{s_B}{c_v^B} k^2 =0.
	\end{equation}
	If we perform exactly the same calculation for Israel-Stewart we have
	\begin{equation}
		\delta T^{00} = \delta e_I   \quad \quad \quad  \delta T^{01} =(e_I  + p_I) \delta u^1  \quad \quad \quad  \delta T^{11} = \delta p_I +\delta \Pi \, ,
	\end{equation}
	so that $\partial_\mu T^{\mu \nu}=0$ becomes
	\begin{equation}
		-i\omega \delta e_I + ik (e_I  + p_I) \delta u^1 =0  \quad \quad \quad -i\omega (e_I  + p_I) \delta u^1 + ik (\delta p_I + \delta \Pi) =0,
	\end{equation}
	while \eqref{IsraelRelax} is
	\begin{equation}
		-i\omega \tau_\Pi \delta \Pi  + \delta \Pi = -ik \zeta \delta u^1.
	\end{equation}
	We, therefore, obtain the condition
	\begin{equation}\label{SpectrumIS}
		\tau_\Pi \, \omega^3 + i \omega^2 -\bigg(\frac{\zeta}{e_I+p_I} + \frac{s_I}{c_v^I} \tau_\Pi \bigg) \, k^2 \omega - i\frac{s_I}{c_v^I} k^2 =0.
	\end{equation}
	Comparing \eqref{spectrumBDNK} with \eqref{SpectrumIS} we see that the spectra of the two theories coincide (to the leading order) provided that \eqref{matchingCoefficients} holds.

\begin{figure}[ht]
          		\includegraphics[width=0.95\textwidth]{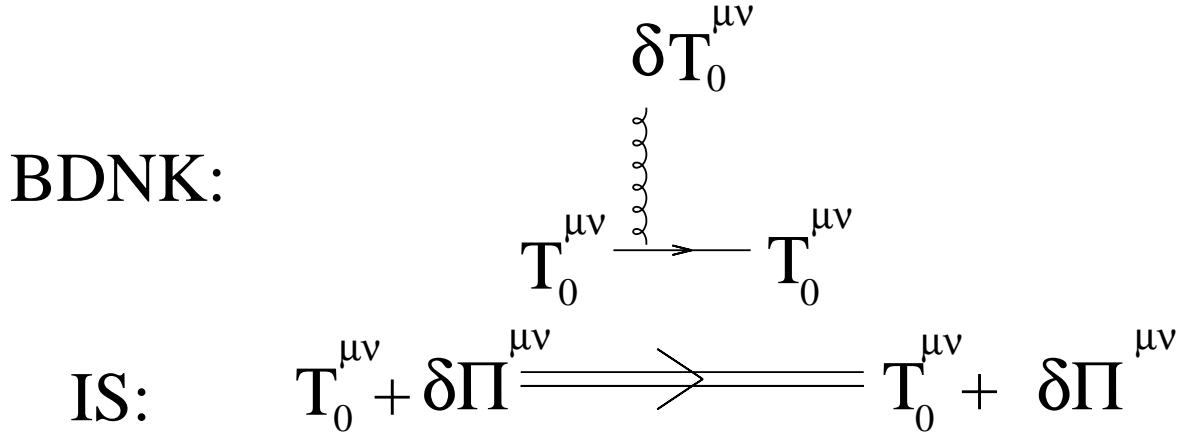}
		\caption{\label{feyndiag}A graphical representation of the hydrodynamic sound wave propagator in different representations.  In one case (upper figure), it is a sum of a thermal fluctuation and the non-hydrodynamic mode. In the second, the latter is reabsorbed into a redefinition of the free sound wave propagator}
	\end{figure}

\color{black}        Let us examine what  \eqref{spectrumBDNK} and \eqref{SpectrumIS} mean physically:   In \eqref{SpectrumIS} the $\order{\omega^3,k^2 \omega}$ terms represents the dissipative tensor $\Pi$, while in  \eqref{spectrumBDNK} they are related to the energy density and temperature redefinitions encoded in $\chi$.    
	$c_v$ as defined in the Landau frame also controls thermal energy fluctuations with
$$\ave{(\Delta e)^2}/\ave{ e}^2 \sim c_v/T^3.$$
This means that, for a perturbation of a given wavenumber and frequency, the ``IS observer'' will interpret $\order{\omega^3,k^2 \omega}$ terms as deviations from local equilibrium whose energy will be transformed into heat as the system relaxes (\figref{feyndiag} (a)).   The ``BDNK observer'' will see a thermal fluctuation interacting with the sound wave, with the interaction converting some of the sound wave's energy into heat (\figref{feyndiag} (b)).   However, since the ``BDNK observer'' will consider the thermal fluctuation as a microstate, the {\em macrostate physics} seen by both observers will be the same.   The different pictures  reflect uncertainty in the rest frame's characteristic thermal volume element, used for ($\sim 1/T$) hydrodynamic coarse-graining.   IS and BDNK volume elements reflect different choices compatible with this uncertainty.\color{black}

	In the next subsection, we will obtain a way to go beyond this first order, but it is clear that once back-reaction is included the BDNK propagator will become less simple.
	\subsection{Bulk viscosity and reparametrization \label{bulkrep}}
	For the case of bulk viscosity, the transformations are a Lorentz transformation to the rest frame w.r.t. $u_\mu$, followed by dilatation and a reverse Lorentz transformation
	\begin{equation}
		\label{transform}
		dM_{\alpha \beta} = (\Lambda_{\alpha}^\mu)^{-1} dU_{\mu \nu} (\Lambda_{\beta}^\nu) \eqcomma dU_{\mu \nu} = g_{\mu \nu} d\mathcal{A}(x)\,.
	\end{equation}
	Moreover, \eqref{avconst} can be trivially satisfied when one looks at \eqref{BridgeFormulaAA}. If a $d\mathcal{A}$ is found in one cell, the rest of the fluid will obey the equation
	\begin{equation}
		\label{alphaconst}
		\delta\Pi(x^\mu + dx^\mu) = -dx^\mu \partial_\mu \mathcal{A} \eqcomma
		\partial_\mu \mathcal{A} =  -\frac{d}{dT} \left(\frac{s}{c_V} \right) \partial_\mu T\,.
	\end{equation}
        \color{black}The substition of \eqref{identifications} is equivalent to integrating \eqref{alphaconst} in every cell from 0 $\Pi$.   A partial integration is equivalent to a hybrid anisotropic Israel-Stewart setup of the type of \cite{spal}
        In the Zubarev picture, a fluid with just bulk viscosity would require a foliation tracking expansion
        \begin{equation}
\frac{\partial \Sigma^\mu}{\partial x_{\nu}} \sim \frac{1}{2} \left(\partial^\mu u^\nu + \partial^\nu u^\mu \right)
        \end{equation}
        for the co-moving bulk dissipative term $\Pi$, since the foliation is co-moving w.r.t. the compression or dilatation.   A $\mathcal{A}\ne 0$ in \eqref{alphaconst} acting locally on $d\Sigma^\mu$ will alter the value of the effective $\pi$ as well as the gradient of $u_\mu=\beta_\mu/(-\beta_\nu \beta^\nu)$ according to \eqref{msigmadef}.
        \color{black}.
        
	In the leading order in hydrodynamic fluctuations, this prescription is equivalent to deforming the Boltzmann equation formula for $\zeta$  \cite{jeonb}
	\begin{equation}
		\label{bulkjeon}
		\zeta=\frac{1}{T}\int \frac{d^3 k}{(2\pi)^2 E_k} \left( \frac{k^2}{3}-v_s^2 (k^2+m^2) \right) n(E_k)\left(1+n(E_k)\right)A(k) \eqcomma 
	\end{equation}
	where $A(k)$ is the bulk kernel using \eqref{transform} as well as $c_V$ and $s$ from $\lnz$.  However, as argued earlier, the Boltzmann equation approximation is likely to fail in the regime of applicability of this theory \color{black} and a Vlasov component, as per \eqref{vlasovfunc}, will be necessary \color{black}.
		
	The operation of the bulk viscosity is equivalent to a conformal transformation, 
	\begin{equation}
		\label{confshift}
		\mathcal{L}  \rightarrow \mathcal{L}+T_{\mu}^{\mu} u_\alpha \partial^\alpha \mathcal{A}  \end{equation}
	Hence, one can use the bulk-viscosity  Lagrangian derived in~\cite{grozdanov,montenegro}, to explicitly work out the constraint in a bulk viscosity only first order theory functional integral. From~\cite{montenegro}, the Lagrangian of this theory is formulated via doubled variables, with equilibrium degrees of freedom $B$ and $K_\mu$ (respectively the square of the entropy and it's current) and the non-equilibrium $\Pi$.   The doubled variable dissipative Lagrangian (see appendix of \cite{montenegro} and references therein) is
	\begin{equation}
		\label{isl}
		\mathcal{L} = \mathcal{L}_{\rm ideal}(B_{\pm}) + \mathcal{L}_{\rm IS-bulk}\,,
	\end{equation}
	with
	\[\ \mathcal{L}_{\rm IS-bulk}= \frac{1}{2}\tau_\pi^\zeta \left(\Pi_- u^\alpha_+ \partial_\alpha \Pi_{ +} -  \Pi_+ u^\alpha_- 
	\partial_\alpha \Pi_{ -}\right)  + \frac{1}{2} \Pi_{\pm}^2  
	+ \underbrace{\left[  K_\mu \partial^\mu B \right]_{\pm}}_{\sim \sigma^\zeta}\,,
	\]
	and apply the transformation (\ref{transform}) and (\ref{alphaconst}) as a constraint.
	This will have to be implemented via the whole functional integral:
	\begin{align}
		\int \mathcal{D}B_{\pm} \mathcal{D}\Pi_{\pm} \rightarrow \int \mathcal{D}\mathcal{A} \mathcal{D}B_{\pm} \mathcal{D}\Pi_{\pm}  \delta\left( \partial_\mu \mathcal{A}  -\frac{d}{dK_\mu K^\mu} \left(\frac{s}{c_V} \right) \partial_\mu (K_\mu K^\mu) \right) 
	\end{align}
	this is an illustration of how first order theories require transformations that appear as constraints in the functional integral. Such theories therefore can not be reduced to pure Lagrangian form.
	
	As a final comment, we remark that the discussion in this section is inherently relativistic because \eqref{confshift} is specific to the Lorentz group.   Non-relativistic conformal transformations are added to the Schrodinger group rather than the Lorentz one, and the Kubo formulae for bulk viscosity are different~\cite{sonconf}.
	\section{Functional methods and micro-macro correlations} \label{functional}
	The previous sections introduced a class of spacetime transformations that, close to local equilibrium (defined by the approximate validity of \eqref{crooks}, as opposed to a gradient expansion), should leave hydrodynamics invariant. Note that \eqref{crooks} may still hold, even when gradients are large. At first, the apparent non-inertial nature of these transformations generates confusion as to the effective theory: Where are the Christoffel symbols associated with the pressure forces?   As the relationship with the Zubarev statistical operator makes clear, as long as one only looks at the coarse-grained variables, there is no need for these:  The non-inertial transformations are not physical, but rather sample the possible fluctuations in energy-momentum compatible with the observed $T_{\mu \nu}$.   The situation changes if one wants to build hydrodynamics ``top-down'', including the microscopic as well as the macroscopic fluctuations.
	
	Within the approach of~\cite{grossi} the effect of both classes of fluctuations can be combined by defining an effective action $\Gamma$, within the microscopic fields $\phi$ as a Legendre transformation on $\lnz$
	\begin{equation}
		\Gamma(\phi)={\rm Sup_J}\left( \int J(x)\phi(x)- i \lnz[J] \right)\,,
	\end{equation}
	while applying the dilatation operator on $\phi(x)$.
	
	In particular, applying the transformation to linear transformations will
	go from an equation of the form  \eqref{SpectrumIS} to
	\eqref{spectrumBDNK}.   Any perturbation in this flow should not be considered as a physical perturbation (altering the entropy of the system) but rather will be a {\em reparametrization}, moving energy density to and from the ``equilibrium'' part.
	In this vein, let us ``start'' with the Landau condition, \eqref{eq:landau}, and change coordinates using the Christoffel symbol defined from \eqref{element1},\eqref{element2}
	\begin{equation}
		\label{liegroup}
		- \partial_\mu M_{\alpha \beta}  = \Gamma_{\alpha \gamma \mu} M^{\gamma}_{\beta} + \Gamma_{\beta \gamma \mu} M^{\gamma}_{\alpha} \eqcomma \Gamma_{\beta \gamma \mu}=\frac{1}{2} \left( \partial_\gamma M_{\alpha \beta}+\partial_\beta M_{\beta \alpha }-\partial_\alpha M_{ \gamma \beta} \right)
	\end{equation}
	It is nevertheless what mathematicians call a ``passive transformation'' since observables are unchanged.
	
	For fluids with flow gradients, transformations of this type are not inertial since $M_{\alpha \beta}$ change in time. Thus they do not correspond to conserved charges but rather the sort of extended symmetries, corresponding to non-conserved Noether currents, in~\cite{grossi}.  As long as they represent
	flow in the space of \eqref{renoflow}, they do represent conserved currents.  
	
	This gives us a recipe to incorporate the symmetry group into the generating functional approach of~\cite{grossi}. One can actually perform a Legendre transform of \eqref{lnzeq} enlarged with an auxiliary field $M_{\alpha \beta}$\\
	\begin{equation}
		\label{lagmult}
		\mathcal{Z}[J_{\alpha \beta \gamma}] = \int \mathcal{D}\phi \mathcal{D}M_{\alpha \beta} \exp\left[ \int det[M] d^4 x \mathcal{L}\left( \phi,\partial_\mu + \Gamma...\right)+ \int d\Sigma^\gamma M^{\alpha \beta} J_{\alpha \beta \gamma} \right]
	\end{equation}
	Note that $S(\phi)$ is the microscopic action, including all microscopic fluctuations.  
	Provided the foliation is not changed (we remain within the same point in the flow defined in \eqref{renoflow} during the fluctuating fluid's evolution)
	this action has the conserved currents
	\begin{equation}
		\label{conscharge}
		J_{\mu \nu \gamma} =\frac{1}{\sqrt{g}}\frac{\delta \lnz [J']}{\delta \Gamma^{\alpha \nu \gamma}} \eqcomma
		D_\mu J^{\mu \nu \gamma}=0
	\end{equation}
	For the bulk viscosity case, \eqref{conscharge} reduces to a Weyl transformation (See Sec.~4.4 of~\cite{grossi}), $J_{\mu \nu \gamma} \sim \partial_{\mu}\mathcal{A}(x) g_{\nu \gamma}+c.p.$, as is apparent from \eqref{confshift}.
	
	Equations (\ref{lagmult}) and (\ref{conscharge}) represent the generalization of the Legendre transformation between energy and temperature in the Grand Canonical and Canonical ensembles respectively, from a hydrostatic limit to an approximately locally equilibrated hydrodynamic one.
	Just like there is an uncertainty in temperature once energy is fixed, there should be a fluctuation flow, say $\delta\beta_\mu$, when momenta currents are fixed. A range of flow choices are permissible with the non-inertial transformations between them being regulated by $\Gamma_{\alpha \beta \mu}$.
	The expressions describing uncertainties are much more complicated since both mean free path and fluctuations enter them.
	
	Mathematically, $\Gamma_{\alpha \beta \mu}$ can be related to pseudo-gauge transformations since these can be thought of as representing a generalization of angular momenta~\cite{brauner} involving local simultaneous field redefinitions and boosts. 
	\color{black}
	The exact partition function is invariant under field redefinitions \cite{peskin} because of the functional integral and coordinate transformations because of covariance.   However, any operator truncation will break this invariance to the degree that fluctuations and correlations are integrated out, and indeed, as worked out for instance ~\cite{fluctgauge}, coarse-graining generally breaks such pseudo-gauge symmetries.   The symmetries discussed here, when included in the coarse-graining, could make way for an explicitly pseudogauge invariant version of fluid dynamics, in the case of spin with torsional components ~\cite{grossi}\color{black}.
	
	The choice of $J_{\mu \nu \gamma}$ which minimizes
	\begin{equation}
		\label{omega2}
		\Psi^2= \int d^4 x \left( \Pi_{\mu \nu} - \sigma_{\mu \nu} \right)^2\,,
	\end{equation}
	with
	\[\ \sigma^{\mu \nu} =\eta'(\alpha_I,\beta_I,\mathcal{A})  \Delta^{\mu \beta} \Delta^{\nu \alpha} \left(\partial_\alpha u_\beta+\partial_\beta u_\alpha-\frac{2}{3} g_{\alpha \beta} \partial _\gamma u^\gamma \right) +\Delta^{\mu \nu} \zeta'(\alpha_I,\beta_I,\mathcal{A}) \partial_\gamma u^\gamma \]
	as in \eqref{omega2} is ``as close'' as the system gets to BDNK.  
	Note that $\eta',\zeta'$ flow according to \eqref{renoflow} and \eqref{constraint}.  
	Generally, \eqref{conscharge} should be related to the ways we can choose ``divergence-type'' terms of \cite{geroch}, and \eqref{renoflow} and \ref{omega2} can be used to define a flow within this space.
	
	To investigate the stability, one could look at a generalized entropy current as in~\cite{gavassino}. The number of physical microstates should be independent of gauge parametrizations. Thus 
	\begin{equation}
		\label{gaugeen}
		s^\mu \rightarrow s^\mu(e,u_\mu,\Pi_{\mu \nu},T_{\mu \nu}) \equiv  s_{0}^\mu (e,u_\mu,\Pi_{\mu \nu}) -\beta_i(x)T_{\alpha \nu} \Gamma^{\mu \alpha \nu}_i \eqcomma  \beta_i(x) \Gamma^{\mu \alpha \nu}_i = \lambda_i(x) J^{\mu \nu} u^\nu(x)\,,
	\end{equation}
	with $\lambda_i$ is chosen to be invariant under \eqref{liegroup}.
	This is a way to calculate the {\em microscopic entropy} corresponding to the Gibbian entropy discussion in Sec.~\ref{reptransport}. 
	In such an equation $n^\mu,\Pi_{\mu \nu},T$ and $u_\mu$ are gauge dependent, but $s$ is not (the coarse-grained $s$, examined in the previous sections, generally is gauge dependent).  The positivity, as well as the relation between entropy non-decrease, stability \cite{gavassino,gavagibbs}, and causality~\cite{GavaCausality2021}, could be related to the allowed gauge space.
	
	Extending thermodynamic Bayesian uncertainties to flows could help to clarify the ``unreasonable effectiveness of hydrodynamics''~\cite{unreasonable} without recourse to hypotheses, such as ``hydrodynamization'' and ``hydrodynamic attractors''~\cite{attractors} which in their current status strongly depend on the system's symmetries.
	It has long been known that Liouville's theorem means that a chaotic system with comparatively few degrees of freedom looks ``ergodic'' when only "elementary" observables (1-particle expectation values and 2-particle correlators) are measured.  Within quantum mechanics, this is known as ``canonical typicality''~\cite{typicality,typicality1}, and can drastically reduce thermalization times (see the discussion on the Berry conjecture in \cite{berry}).
	The ideas developed here can be used to generalize these concepts to a flowing fluid. If it is strongly coupled (``locally chaotic'' enough), and one measures only conserved coarse-grained quantities, then there will be a range of ways to describe this system.  Some choices for flow will correspond to a path in phase space that mimics that of an ideal fluid.   What makes this picture compelling for small systems~\cite{zajc} is that the range of allowed choices is partially determined by local thermodynamic fluctuations, and hence increases with particle number.   Provided the system is strongly interacting enough, therefore, it is not surprising small systems are described by hydrodynamics, but the underlying picture is in a sense opposite to that of attractors: Hydrodynamics is one within an ensemble of choices;  As long as "near-equilibrium", defined by the applicability of \eqref{crooks} holds (fluctuations are mainly determined by microstates), frame choices obeying nearly ideal hydro will be compatible
with the evolution of the system at the fluctuation scale. \color{black}  This requirement could in principle be much more easy to reach than that of a small Knudsen number, since a mixture of coherent (one microstate) and incoherent (many microstates) systems could have approximate validity of \eqref{crooks}, as has been known for a long time in the literature dealing with mesoscopic systems~\cite{landi}.   This might be a relativistic equivalent of the very well known ``Brazil nut effect'' where a comparatively small number of extented correlated particles (nuts of various shapes, corn-flakes) subjected to random vibrations exhibit phenomena (convective flow, buoyancy) typical for a fluid \cite{brazilnut}.   Clearly such a system is well outside of applicability for transport theory, motivating the idea that hydrodynamics could emerge in systems of comparatively small numbers ofl strongly interacting particles.   In such a system Eigenstate thermalization \cite{typicality1} in every cell could provide the equivalent of the "shaking" in the Brazil nut effect, while the paucity of degrees of freedom in each cell provides for multiple gauge choices, each with different Boltzmann entropy but in the same Gibbs entropy class.   The lowest Boltzmann entropy corresponds to the ``most hydrodynamic`` description\color{black}
	
	The intuition that the microscopic fluctuating scale functions as a lower limit on dissipation, developed in \cite{stick,ryblewski} among others, would need to be drastically revisted, because, in a manner reminiscent of the Gauge potential, the equilibrium and non-equilibrium parts of $T_{\mu \nu}$ are ambiguus and this {\em reduces} the role of fluctuations in bringing the system away from local equilibrium.
	\section{Concluding remarks}\label{sec:conclusion}
	In conclusion, we have made the argument that different frame choices required to define a hydrodynamic evolution, for example, IS and BDNK, reflect ambiguity in connecting observables, correlations of $T_{\mu \nu}$ to the underlying dynamics around local equilibrium, determined both by microscopic thermodynamic fluctuations and dissipative hydrodynamics. We have illustrated this by calculating the bulk viscosity matching within both the microscopic definition in terms of microscopic quantities and mesoscopic physics in terms of sound waves.
	This opens the road to a detailed accounting of causality and well-posedness in terms of fluctuating quantities.   \color{black}It also suggests a novel explanation, in a sense opposite to the ``attractor`` paradigm, for the seeming applicability of hydrodynamics to systems with a comparatively small number of particles, by illustrating that such strongly interacting fluctuating systems are open to a variety of equivalent descriptions, some of which look hydrodynamical.
We hope to develop this picture into a quantitative model and relate it to microscopic theories in forthcoming works. \color{black}
	
	
	L.G.~acknowledges support from the Polish National Science Centre grants OPUS 2019/33/B/ST9/00942. G.T.~acknowledges support from Bolsa de produtividade CNPQ 306152/2020-7, Bolsa de pesquisa FAPESP 2021/01700-2 and Partecipation to Tematico FAPESP, 2017/05685-2. T.D. is supported by the US-DOE Nuclear Science Grant No. DE-SC0020633.	 M.S. acknowledges support by the Deutsche Forschungsgemeinschaft (DFG, German Research Foundation) through the CRC-TR 211 'Strong-interaction matter under extreme conditions'– project number 315477589 – TRR 211.
\appendix
\section{Linearization and entropy in an effective theory \label{apptoy}}  
Consider a $0+1$ dimensional model with degrees of freedom $u,q$ (approximating velocity and heat) and entropy $s$, related by
 \begin{equation}
s=-u^2+u q - q^2
\end{equation}
and exact evolution equations
\begin{equation}
\label{evolqmu}
\dot{u}=-\frac{2 u+q}{3} \eqcomma \dot{q}=-\frac{2 q+u}{3}
\end{equation}.
It is clear that the exact evolution equation for entropy is
\begin{equation}
\dot{s}=u^2 + q^2 \geq 0
\end{equation} 
so the "H-theorem" holds.
If one however linearizes in q, $s\simeq -u^2+u q$ the entropy evolution equation becomes 
\begin{equation}
\label{slin}
\left. \dot{s} \right|_{linear}=u^2 -\frac{2}{3} u q -\frac{1}{3} q^2
\end{equation} 
while \eqref{evolqmu} contunes invariant, since it is exact.    This means that if $u$ is small, entropy can actually "decrease" in \eqref{slin}, but only because once $\dot{s}$ is computed from the {\em linearized} definition of $s$ and the exact equations of motion the H theorem is lost.   

We argue that approaches such as \cite{kovtunani} are similar:  The evolution of $f(x,p)$ follows closely the
exact Boltzmann equation, as shown in \cite{denicol}, but the definition of entropy as $f \ln f$ receives large corrections and this leads to the fact that while entropy is conserved in the exact Boltzmann equation, it is not in the truncated one.

Does this mean that the truncated equation is a good approximation of the Boltzmann equation?  Perhaps, but one must be careful in using the truncation to derive hydrodynamics because the resulting transport coefficients can not anymore be derived in terms of the infrared limit of the fluctuation-dissipation theorem, since entropy determines fluctuations.      Which implies that corrections {\em to the Boltzmann equation} from deviations of ensemble averaging and molecular chaos are likely to be of the same order as differences between truncated hydrodynamics (which occasionally violates the H-theorem) and the full Boltzmann equation (which never does).
A consistent treatment of fluctuation and dissipation, as advocated in this paper, would then be needed, perhaps based on Vlasov terms and functionals as in \eqref{vlasovfunc}.
		\medskip
\bibliographystyle{unsrt}
	\bibliography{main}

\begin{thebibliography}{10}

\bibitem{zajc}
James~L. Nagle and William~A. Zajc.
\newblock {Small System Collectivity in Relativistic Hadronic and Nuclear
  Collisions}.
\newblock {\em Ann. Rev. Nucl. Part. Sci.}, 68:211--235, 2018.

\bibitem{kodama}
R.~Derradi~de Souza, Tomoi Koide, and Takeshi Kodama.
\newblock {Hydrodynamic Approaches in Relativistic Heavy Ion Reactions}.
\newblock {\em Prog. Part. Nucl. Phys.}, 86:35--85, 2016.

\bibitem{crooks}
Giorgio Torrieri.
\newblock {Fluctuating Relativistic hydrodynamics from Crooks theorem}.
\newblock {\em JHEP}, 02:175, 2021.

\bibitem{nicolis}
Solomon Endlich, Alberto Nicolis, Riccardo Rattazzi, and Junpu Wang.
\newblock {The Quantum mechanics of perfect fluids}.
\newblock {\em JHEP}, 04:102, 2011.

\bibitem{nicolis2}
S.~Dubovsky, T.~Gregoire, A.~Nicolis, and R.~Rattazzi.
\newblock {Null energy condition and superluminal propagation}.
\newblock {\em JHEP}, 03:025, 2006.

\bibitem{beta}
F.~Becattini, L.~Bucciantini, E.~Grossi, and L.~Tinti.
\newblock {Local thermodynamical equilibrium and the beta frame for a quantum
  relativistic fluid}.
\newblock {\em Eur. Phys. J. C}, 75(5):191, 2015.

\bibitem{jackiw}
R.~Jackiw, V.~P. Nair, S.~Y. Pi, and A.~P. Polychronakos.
\newblock {Perfect fluid theory and its extensions}.
\newblock {\em J. Phys. A}, 37:R327--R432, 2004.

\bibitem{glorioso}
Hong Liu and Paolo Glorioso.
\newblock {Lectures on non-equilibrium effective field theories and fluctuating
  hydrodynamics}.
\newblock {\em PoS}, TASI2017:008, 2018.

\bibitem{grozdanov}
Sa\v{s}o Grozdanov and Janos Polonyi.
\newblock {Viscosity and dissipative hydrodynamics from effective field
  theory}.
\newblock {\em Phys. Rev. D}, 91(10):105031, 2015.

\bibitem{ryblewski}
David Montenegro, Radoslaw Ryblewski, and Giorgio Torrieri.
\newblock {Relativistic fluid dynamics and its extensions as an effective field
  theory}.
\newblock {\em Acta Phys. Polon. B}, 50:1275, 2019.

\bibitem{montenegro}
David Montenegro and Giorgio Torrieri.
\newblock {Causality and dissipation in relativistic polarizable fluids}.
\newblock {\em Phys. Rev. D}, 100(5):056011, 2019.

\bibitem{landau}
L.~D. Landau and E.~M. Lifshitz.
\newblock {\em Fluid Mechanics}.
\newblock Butterworth-Heinemann, 2 edition, January 1987.

\bibitem{Eckart}
Carl Eckart.
\newblock {The Thermodynamics of irreversible processes. 3.. Relativistic
  theory of the simple fluid}.
\newblock {\em Phys. Rev.}, 58:919--924, 1940.

\bibitem{kovfluct}
Pavel Kovtun.
\newblock {Lectures on hydrodynamic fluctuations in relativistic theories}.
\newblock {\em J. Phys. A}, 45:473001, 2012.

\bibitem{hislim}
W.~A. Hiscock and L.~Lindblom.
\newblock {Stability and causality in dissipative relativistic fluids}.
\newblock {\em Annals Phys.}, 151:466--496, 1983.

\bibitem{sc-problem-85}
William~A. Hiscock and Lee Lindblom.
\newblock {Generic instabilities in first-order dissipative relativistic fluid
  theories}.
\newblock {\em Phys. Rev. D}, 31:725--733, 1985.

\bibitem{sc-problem-87}
William~A. Hiscock and Lee Lindblom.
\newblock {Linear plane waves in dissipative relativistic fluids}.
\newblock {\em Phys. Rev. D}, 35:3723--3732, 1987.

\bibitem{jeon}
Alina Czajka and Sangyong Jeon.
\newblock {Kubo formulas for the shear and bulk viscosity relaxation times and
  the scalar field theory shear $\tau_\pi$ calculation}.
\newblock {\em Phys. Rev. C}, 95(6):064906, 2017.

\bibitem{IsraelStewart}
W.~Israel and J.~M. Stewart.
\newblock {Transient relativistic thermodynamics and kinetic theory}.
\newblock {\em Annals Phys.}, 118:341--372, 1979.

\bibitem{janik}
Michal~P. Heller, Romuald~A. Janik, Micha\l{} Spali\'nski, and Przemys\l{}aw
  Witaszczyk.
\newblock {Coupling hydrodynamics to nonequilibrium degrees of freedom in
  strongly interacting quark-gluon plasma}.
\newblock {\em Phys. Rev. Lett.}, 113(26):261601, 2014.

\bibitem{trachenko}
K~Trachenko and V~V Brazhkin.
\newblock Collective modes and thermodynamics of the liquid state.
\newblock {\em Reports on Progress in Physics}, 79(1):016502, Dec 2015.

\bibitem{Bemfica:2017wps}
F\'abio~S. Bemfica, Marcelo~M. Disconzi, and Jorge Noronha.
\newblock {Causality and existence of solutions of relativistic viscous fluid
  dynamics with gravity}.
\newblock {\em Phys. Rev. D}, 98(10):104064, 2018.

\bibitem{kovtun}
Pavel Kovtun.
\newblock {First-order relativistic hydrodynamics is stable}.
\newblock {\em JHEP}, 10:034, 2019.

\bibitem{kovtun2}
Raphael~E. Hoult and Pavel Kovtun.
\newblock {Stable and causal relativistic Navier-Stokes equations}.
\newblock {\em JHEP}, 06:067, 2020.

\bibitem{bdnk}
F\'abio~S. Bemfica, Marcelo~M. Disconzi, and Jorge Noronha.
\newblock {Nonlinear Causality of General First-Order Relativistic Viscous
  Hydrodynamics}.
\newblock {\em Phys. Rev. D}, 100(10):104020, 2019.

\bibitem{Bemfica:2020zjp}
Fabio~S. Bemfica, Marcelo~M. Disconzi, and Jorge Noronha.
\newblock {General-Relativistic Viscous Fluid Dynamics}.
\newblock 9 2020.

\bibitem{spal}
Jorge Noronha, Micha\l{} Spali\'nski, and Enrico Speranza.
\newblock {Transient Relativistic Fluid Dynamics in a General Hydrodynamic
  Frame}.
\newblock 5 2021.

\bibitem{yarom}
Jyotirmoy Bhattacharya, Sayantani Bhattacharyya, Shiraz Minwalla, and Amos
  Yarom.
\newblock {A Theory of first order dissipative superfluid dynamics}.
\newblock {\em JHEP}, 05:147, 2014.

\bibitem{biro}
P.~Van and T.~S. Biro.
\newblock {First order and stable relativistic dissipative hydrodynamics}.
\newblock {\em Phys. Lett. B}, 709:106--110, 2012.

\bibitem{biro2}
P.~Van and T.~S. Biro.
\newblock {Relativistic hydrodynamics - causality and stability}.
\newblock {\em Eur. Phys. J. ST}, 155:201--212, 2008.

\bibitem{bantilan}
Hans Bantilan, Yago Bea, and Pau Figueras.
\newblock {Evolutions in first-order viscous hydrodynamics}.
\newblock 1 2022.

\bibitem{gavassino}
Lorenzo Gavassino, Marco Antonelli, and Brynmor Haskell.
\newblock {When the entropy has no maximum: a new perspective on the
  instability of the first-order theories of dissipation}.
\newblock {\em Phys. Rev. D}, 102(4):043018, 2020.

\bibitem{ShokryTaghi2020}
M.~{Shokri} and F.~{Taghinavaz}.
\newblock {Conformal Bjorken flow in the general frame and its attractor:
  Similarities and discrepancies with the M{\"u}ller-Israel-Stewart formalism}.
\newblock {\em arXiv e-prints}, page arXiv:2002.04719, February 2020.

\bibitem{GavaUeit2021}
Lorenzo {Gavassino} and Marco {Antonelli}.
\newblock {Unified Extended Irreversible Thermodynamics and the stability of
  relativistic theories for dissipation}.
\newblock {\em Frontiers in Astronomy and Space Sciences}, 8:92, June 2021.

\bibitem{jaynes}
E.~T. Jaynes.
\newblock Gibbs vs boltzmann entropies.
\newblock 33:391--398, 1965.

\bibitem{brazilnut}
James~B. Knight, E.~E. Ehrichs, Vadim~Yu. Kuperman, Janna~K. Flint, Heinrich~M.
  Jaeger, and Sidney~R. Nagel.
\newblock Experimental study of granular convection.
\newblock {\em Phys. Rev. E}, 54:5726--5738, Nov 1996.

\bibitem{LIU1986191}
Relativistic thermodynamics of gases.
\newblock {\em Annals of Physics}, 169(1):191--219, 1986.

\bibitem{geroch}
Robert~P. Geroch and L.~Lindblom.
\newblock {Dissipative relativistic fluid theories of divergence type}.
\newblock {\em Phys. Rev. D}, 41:1855, 1990.

\bibitem{Aguilar:2017ios}
Milton Aguilar and Esteban Calzetta.
\newblock {Causal Relativistic Hydrodynamics of Conformal Fermi-Dirac Gases}.
\newblock {\em Phys. Rev. D}, 95(7):076022, 2017.

\bibitem{Miron-Granese:2020mbf}
Nahuel Mir\'on-Granese, Alejandra Kandus, and Esteban Calzetta.
\newblock {Nonlinear Fluctuations in Relativistic Causal Fluids}.
\newblock {\em JHEP}, 07:064, 2020.

\bibitem{Calzetta:1997aj}
Esteban Calzetta.
\newblock {Relativistic fluctuating hydrodynamics}.
\newblock {\em Class. Quant. Grav.}, 15:653--667, 1998.

\bibitem{grossi}
Stefan Floerchinger and Eduardo Grossi.
\newblock {Conserved and non-conserved Noether currents from the quantum
  effective action}.
\newblock 2 2021.

\bibitem{kunihiro}
K.~Tsumura, T.~Kunihiro, and K.~Ohnishi.
\newblock {Derivation of covariant dissipative fluid dynamics in the
  renormalization-group method}.
\newblock {\em Phys. Lett. B}, 646:134--140, 2007.
\newblock [Erratum: Phys.Lett.B 656, 274 (2007)].

\bibitem{kunihiro2}
Kyosuke Tsumura and Teiji Kunihiro.
\newblock {First-Principle Derivation of Stable First-Order Generic-Frame
  Relativistic Dissipative Hydrodynamic Equations from Kinetic Theory by
  Renormalization-Group Method}.
\newblock {\em Prog. Theor. Phys.}, 126:761--809, 2011.

\bibitem{kovtunani}
Raphael~E. Hoult and Pavel Kovtun.
\newblock {Causal first-order hydrodynamics from kinetic theory and
  holography}.
\newblock 12 2021.

\bibitem{denicol}
Gabriel~S. Rocha, Gabriel~S. Denicol, and Jorge Noronha.
\newblock {Novel Relaxation Time Approximation to the Relativistic Boltzmann
  Equation}.
\newblock {\em Phys. Rev. Lett.}, 127(4):042301, 2021.

\bibitem{lanford}
Oscar~E. {Lanford}.
\newblock {The hard sphere gas in the Boltzmann-Grad limit}.
\newblock {\em Physica A Statistical Mechanics and its Applications},
  106(1-2):70--76, March 1981.

\bibitem{kadanoff}
Leo~P. Kadanoff and Paul~C. Martin.
\newblock Hydrodynamic equations and correlation functions.
\newblock 24:419--469, 1963.

\bibitem{gale}
J.~I. Kapusta and Charles Gale.
\newblock {\em {Finite-temperature field theory: Principles and applications}}.
\newblock Cambridge Monographs on Mathematical Physics. Cambridge University
  Press, 2011.

\bibitem{huang}
Kerson Huang.
\newblock {\em Statistical Mechanics}.
\newblock John Wiley \& Sons, 2 edition, 1987.

\bibitem{ensemble1}
Eric~M. Pearson, Timur Halicioglu, and William~A. Tiller.
\newblock {Laplace-transform technique for deriving thermodynamic equations
  from the classical microcanonical ensemble}.
\newblock {\em Phys. Rev. A}, 32:3030--3039, 1985.

\bibitem{ensemble2}
R.~Hagedorn and K.~Redlich.
\newblock {Statistical Thermodynamics in Relativistic Particle and Ion Physics:
  Canonical or Grand Canonical?}
\newblock {\em Z. Phys. C}, 27:541, 1985.

\bibitem{ensemble3}
Marcus V.~S. Bonanca.
\newblock Fluctuation-dissipation theorem for the microcanonical ensemble.
\newblock {\em Physical Review E}, 78(3), Sep 2008.

\bibitem{tong}
David Tong.
\newblock Lectures on kinetic theory.
\newblock \url{https://www.damtp.cam.ac.uk/user/tong/kintheory/}.
\newblock (Accessed on 07/13/2021).

\bibitem{jeonb}
Sangyong Jeon.
\newblock {Hydrodynamic transport coefficients in relativistic scalar field
  theory}.
\newblock {\em Phys. Rev. D}, 52:3591--3642, 1995.

\bibitem{sonreview}
Dam~T. Son and Andrei~O. Starinets.
\newblock {Viscosity, Black Holes, and Quantum Field Theory}.
\newblock {\em Ann. Rev. Nucl. Part. Sci.}, 57:95--118, 2007.

\bibitem{lindblom}
Lee Lindblom.
\newblock {The Relaxation effect in dissipative relativistic fluid theories}.
\newblock {\em Annals Phys.}, 247:1, 1996.

\bibitem{laurent}
Gabriel~S. Denicol, Jorge Noronha, Harri Niemi, and Dirk~H. Rischke.
\newblock {Origin of the Relaxation Time in Dissipative Fluid Dynamics}.
\newblock {\em Phys. Rev. D}, 83:074019, 2011.

\bibitem{dirkis}
Arus Harutyunyan, Armen Sedrakian, and Dirk~H. Rischke.
\newblock {Relativistic second-order dissipative hydrodynamics from Zubarev's
  non-equilibrium statistical operator}.
\newblock 10 2021.

\bibitem{arnold}
Peter~Brockway Arnold.
\newblock {Quark-Gluon Plasmas and Thermalization}.
\newblock {\em Int. J. Mod. Phys. E}, 16:2555--2594, 2007.

\bibitem{zeroth}
Bruce~R. Pearson, Tarek~A. Yousef, Nils Erland~L. Haugen, Axel Brandenburg, and
  Per-Age Krogstad.
\newblock {The Zeroth Law of Turbulence: Isotropic Turbulence Simulations
  Revisited}.
\newblock {\em Phys. Rev. E}, 70:056301, 2004.

\bibitem{stick}
Pavel Kovtun, Guy~D. Moore, and Paul Romatschke.
\newblock {The stickiness of sound: An absolute lower limit on viscosity and
  the breakdown of second order relativistic hydrodynamics}.
\newblock {\em Phys. Rev. D}, 84:025006, 2011.

\bibitem{hnatic}
M.~V. Altaisky, M.~Hnatich, and N.~E. Kaputkina.
\newblock Renormalization of viscosity in wavelet-based model of turbulence.
\newblock {\em Physical Review E}, 98(3), Sep 2018.

\bibitem{zubbec}
F.~Becattini, M.~Buzzegoli, and E.~Grossi.
\newblock {Reworking the Zubarev's approach to non-equilibrium quantum
  statistical mechanics}.
\newblock {\em Particles}, 2(2):197--207, 2019.

\bibitem{nishioka}
Tatsuma Nishioka.
\newblock {Entanglement entropy: holography and renormalization group}.
\newblock {\em Rev. Mod. Phys.}, 90(3):035007, 2018.

\bibitem{boulware}
Stanley Deser and D.~Boulware.
\newblock {Stress-Tensor Commutators and Schwinger Terms}.
\newblock {\em J. Math. Phys.}, 8:1468, 1967.

\bibitem{jeon2}
Sangyong Jeon and Ulrich Heinz.
\newblock {Introduction to Hydrodynamics}.
\newblock {\em Int. J. Mod. Phys. E}, 24(10):1530010, 2015.

\bibitem{gavagibbs}
L.~Gavassino.
\newblock {Applying the Gibbs stability criterion to relativistic
  hydrodynamics}.
\newblock 4 2021.

\bibitem{jlreno}
Giovanni Jona-Lasinio.
\newblock Renormalization group and probability theory.
\newblock {\em Physics Reports}, 352(4-6):439--458, 2001.

\bibitem{coveft}
Timothy Cohen, Nathaniel Craig, Xiaochuan Lu, and Dave Sutherland.
\newblock {On-Shell Covariance of Quantum Field Theory Amplitudes}.
\newblock 2 2022.

\bibitem{varnold}
V.Arnold and D.Khesin.
\newblock {\em {Topological methods in fluid dynamics}}.
\newblock Springer. Springer, 2011.

\bibitem{Israel2009}
Werner Israel.
\newblock {\em Relativistic Thermodynamics}, pages 101--113.
\newblock Birkh{\"a}user Basel, Basel, 2009.

\bibitem{peskin}
Michael~E. Peskin and Daniel~V. Schroeder.
\newblock {\em {An Introduction to quantum field theory}}.
\newblock Addison-Wesley, Reading, USA, 1995.

\bibitem{GavaCausality2021}
Lorenzo {Gavassino}, Marco {Antonelli}, and Brynmor {Haskell}.
\newblock {Thermodynamic stability implies causality}.
\newblock {\em arXiv e-prints}, page arXiv:2105.14621, May 2021.

\bibitem{villani}
Cedric~Villani Clement~Mouhot.
\newblock {on landau damping}.
\newblock {\em Acta Math.}, 207(1):29--201, 2011.

\bibitem{crismanuel}
P.~F. Kelly, Q.~Liu, C.~Lucchesi, and C.~Manuel.
\newblock {Classical transport theory and hard thermal loops in the quark -
  gluon plasma}.
\newblock {\em Phys. Rev. D}, 50:4209--4218, 1994.

\bibitem{wignerf}
Stanislaw Mrowczynski and Berndt Muller.
\newblock {Wigner functional approach to quantum field dynamics}.
\newblock {\em Phys. Rev. D}, 50:7542--7552, 1994.

\bibitem{rghydro}
Adrian Koenigstein, Martin~J. Steil, Nicolas Wink, Eduardo Grossi, Jens Braun,
  Michael Buballa, and Dirk~H. Rischke.
\newblock {Numerical fluid dynamics for FRG flow equations: Zero-dimensional
  QFTs as numerical test cases - Part I: The $O(N)$ model}.
\newblock 8 2021.

\bibitem{spohn}
Herbert Spohn.
\newblock {The Critical manifold of the Lorentz-Dirac equation}.
\newblock {\em Europhys. Lett.}, 50:287--292, 2000.

\bibitem{ruderman}
M.~Ruderman.
\newblock {Causes of Sound Faster than Light in Classical Models of Ultradense
  Matter}.
\newblock {\em Phys. Rev.}, 172:1286--1290, 1968.

\bibitem{JouBulk}
Mohamed Zakari and David Jou.
\newblock Equations of state and transport equations in viscous cosmological
  models.
\newblock {\em Phys. Rev. D}, 48:1597--1601, Aug 1993.

\bibitem{sonconf}
Yusuke Nishida and Dam~T. Son.
\newblock {Nonrelativistic conformal field theories}.
\newblock {\em Phys. Rev. D}, 76:086004, 2007.

\bibitem{brauner}
Tom\'a\v{s} Brauner.
\newblock {Noether currents of locally equivalent symmetries}.
\newblock {\em Phys. Scripta}, 95(3):035004, 2020.

\bibitem{fluctgauge}
Arpan Das, Wojciech Florkowski, Radoslaw Ryblewski, and Rajeev Singh.
\newblock {Pseudogauge dependence of quantum fluctuations of the energy in a
  hot relativistic gas of fermions}.
\newblock {\em Phys. Rev. D}, 103(9):L091502, 2021.

\bibitem{unreasonable}
Jacquelyn Noronha-Hostler, Jorge Noronha, and Miklos Gyulassy.
\newblock {The unreasonable effectiveness of hydrodynamics in heavy ion
  collisions}.
\newblock {\em Nucl. Phys. A}, 956:890--893, 2016.

\bibitem{attractors}
Aleksi Kurkela, Wilke van~der Schee, Urs~Achim Wiedemann, and Bin Wu.
\newblock {Early- and Late-Time Behavior of Attractors in Heavy-Ion
  Collisions}.
\newblock {\em Phys. Rev. Lett.}, 124(10):102301, 2020.

\bibitem{typicality}
Sheldon Goldstein, Joel~L. Lebowitz, Roderich Tumulka, and Nino Zanghi.
\newblock {Canonical Typicality}.
\newblock {\em Phys. Rev. Lett.}, 96(5):050403, 2006.

\bibitem{typicality1}
Christian Gogolin and Jens Eisert.
\newblock {Equilibration, thermalisation, and the emergence of statistical
  mechanics in closed quantum systems}.
\newblock {\em Rept. Prog. Phys.}, 79(5):056001, 2016.

\bibitem{berry}
F.~Becattini.
\newblock {An Introduction to the Statistical Hadronization Model}.
\newblock In {\em International School on Quark-Gluon Plasma and Heavy Ion
  Collisions: past, present, future}, 1 2009.

\bibitem{landi}
{Giacomo Guarnieri, Gabriel T. Landi, Stephen R. Clark, John Goold}.
\newblock {Thermodynamics of precision in quantum nonequilibrium steady
  states}.
\newblock 8 2019.

\end{thebibliography}
	
\end{document}